\documentclass[journal]{IEEEtran}
\ifCLASSINFOpdf
\else
\fi
\usepackage{amsmath}
\usepackage{graphicx}
\usepackage{subfigure}
\usepackage{amsmath}
\usepackage{algorithm}
\usepackage{algorithm}
\usepackage{algpseudocode}
\usepackage{cite}
\usepackage{multicol}
\usepackage{multirow}
\usepackage{mathtools}
\usepackage{amssymb}
\usepackage{color}
\usepackage{bm}
\usepackage{indentfirst}
\usepackage{nomencl}
\usepackage{booktabs}
\usepackage{amsthm}
\usepackage{siunitx}
\newtheorem{theorem}{\emph{\textbf{Theorem}}}

\newtheorem{remark}{Remark}
\usepackage[numbers,sort&compress]{natbib}
\usepackage{footnote}
\makesavenoteenv{algorithm}

\usepackage[hyphens]{url}
\usepackage[hidelinks]{hyperref}

\begin{document}
%
\title{Stochastic Optimal Control of HVAC System for  Energy-efficient Buildings}
%
%
%

\author{Yu~Yang,~\IEEEmembership{Student Member,~IEEE,}
        Guoqiang~Hu,~\IEEEmembership{Senior Member,~IEEE,}
        and~Costas~J.~Spanos,~\IEEEmembership{Fellow,~IEEE}
\thanks{This  work  is  supported  by  the  Republic  of  Singapore’s  National  Research  Foundation  through  a  grant  to  the  Berkeley  Education  Alliance  for  Research  in  Singapore
	(BEARS)  for  the  Singapore-Berkeley  Building  Efficiency  and  Sustainability  in  the
	Tropics  (SinBerBEST)  Program.  BEARS  has  been  established  by  the  University  of  California,  Berkeley  as  a  center  for  intellectual  excellence  in  research  and  education  in
	Singapore.}
\thanks{Yu Yang is with SinBerBEST, Berkeley Education 	Alliance for Research in Singapore, Singapore 138602 e-mail: (yu.yang@bears-berkeley.sg).}
\thanks{Guoqiang Hu is with the School 	of Electrical and Electronic Engineering, Nanyang Technological University,
	Singapore, 639798 e-mail: (gqhu@ntu.edu.sg).}
\thanks{Costas J. Spanos is with the Department of Electrical Engineering and 	Computer Sciences, University of California, Berkeley, CA, 94720 USA email: (spanos@berkeley.edu).}
}

\maketitle

\begin{abstract}
The heating, ventilation and air-conditioning (HVAC) system accounts for substantial energy use in buildings, whereas a large group of occupants are still  not  actually feeling comfortable staying inside. This poses the issue of  developing energy-efficient HVAC control, i.e., reduce energy use (cost) while simultaneously enhancing human comfort.  
This paper pursues the objective and  studies the  stochastic optimal HVAC control subject to uncertain thermal demand (i.e., the weather and occupancy etc). Particularly, we involve the elaborate predicted mean vote (PMV) thermal comfort model in the optimization. The problem is computationally challenging due to the non-linear and non-analytical constraints imposed by the system dynamics and PMV model.  We make the following  contributions to address it. \emph{First}, we formulate the problem  as a Markov decision process (MDP) which  is a  desirable modeling technique capable of handling the complexities.  \emph{Second}, we propose a gradient-based learning (GB-L) method for progressively learning a stochastic control policy off-line and store it for on-line execution. \emph{Third}, we prove the learning method’s converge to the optimal policies theoretically, and its performance (i.e., energy cost, thermal comfort and on-line computation) for HVAC control via simulations. The comparisons with the existing model predictive control based relaxation (MPC-R) method which is assumed with accurate future information and supposed to provide the near-optimal bounds, show that though there exists some performance loss in energy cost reduction (i.e., 6.5\%), the proposed method can enable efficient on-line implementation (less than 1 second) and provide high probability of thermal comfort under uncertainties. 
\end{abstract}


\IEEEpeerreviewmaketitle

{\small 
\section*{Nomenclature}
\addcontentsline{toc}{section}{Nomenclature}
\textbf{Notations:}
\begin{IEEEdescription}[\IEEEusemathlabelsep\IEEEsetlabelwidth{$P^{\rm{loss}}_{(i,j)}$, $Q^{\rm{lo}}$}]
\item[$t$] Time index.
\item[$\alpha_{\text{w}}$] The absorption coefficient of walls.
\item[\small{$A_{\text{gs}}$}] The area of glass window [\si{\milli^2}].
\item[\small{$A_{\text{wl}}/A_{\text{wr}}$}] The area of left/right wall [\si{\milli^2}].
\item[\small{$C_{\text{p}}$}] The air specific heat [\si{\joule\per({\kilogram \cdot \kelvin}})].
\item[\small{$C_{\text{w}}$}] The  wall capacity [\si{\joule\per{(\kilogram \cdot \kelvin)}}].
\item[$c_t$] The electricity price  [s\$\si{\per{\kilo\watt}}]. 
\item[\small{$\eta$}] The reciprocal of coefficient of performance (COP)  of  chiller.
\item[\small{$G^{\text{fau}}_t/G^{\text{fcu}}_t$}] The supply  air flow rate of FAU/FCU  [\si{\kilogram\per\second}].
\item[\small{$G^{\text{fcu}, \text{r}}$}] The nominal air flow rate of  FCU [\si{\kilogram\per\second}].
\item[\small{$G^{\text{fau}, \text{r}}$}] The nominal air flow rate of  FAU [\si{\kilogram\per\second}].
\item[\small{$\underline{G}^{\text{fau}}/\underline{G}^{\text{fcu}}$}] The lower bound of damper position for FAU/FCU [\si{\kilogram\per\second}].
\item[\small{$\overline{G}^{\textrm{fau}}/\overline{G}^{\text{fcu}}$}] The upper bound of  damper position for  FAU/FCU [\si{\kilogram\per\second}].
\item[\small{$H^{\text{o}}_t/H^{\text{a}}_t$}] Outdoor/indoor relative humidity [\si{\percent}].
\item[\small{$H^{\text{fau}}_t/H^{\text{fcu}}_t$}] Relative humidity of supply air by FAU/FCU [\si{\percent}].
\item[$h_{\text{gs}}/h_{\text{w}}$] Heat transfer coefficient of window/walls [\si{\joule\per{(\milli^2 \cdot \degreeCelsius)}}]. 
\item[$m^{\text{a}}$] The mass of indoor air [\si{\kilogram}].
\item[$m^{\text{wl}}/m^{\text{wr}}$] The mass of left/right wall [\si{\kilogram}].
\item[\small{$F^{\text{fcu}, \text{r}}$}] The nominal fan power of  FCU [\si{\kilo\watt}].
\item[\small{$F^{\text{fau},  \text{r}}$}] The nominal fan power of  FAU [\si{\kilo\watt}].
\item[$Q^{\text{o}}/H^\text{g}$] The average internal heat/humidity generation rate per occupant [\si{\joule\per\second}].
\item[$Q^{\textrm{d}}$] The average heat generation rate of electrical devices per occupant [\si{\joule\per\second}].
\item[$Q^{\text{w}}_t$] Solar radiation density [\si{\joule\per{\milli^2 \cdot \second}}].
\item[\small{$T^{\text{o}}_t/T^{\text{a}}_t$}] Outdoor/indoor temperature  [\si{\degreeCelsius}].
\item[\small{$T^{\text{wl}}_t/T^{\text{wr}}_t$}] The left/right wall temperature [\si{\degreeCelsius}].
\item[\small{$T^{\text{fau} }_t/T^{\text{fcu}}_t$}] The set-point temperature of FAU/FCU   [\si{\degreeCelsius}].
\item[\small{$\underline{T}^{\text{fau}}/\underline{T}^{\text{fcu}}$}] The lower bound of set-point temperature of FAU/FCU [\si{\degreeCelsius}].
\item[\small{$\overline{T}^{\text{fau}}/\overline{T}^{\text{fcu}}$}] The upper bound of set-point temperature of FAU/FCU [\si{\degreeCelsius}].
\end{IEEEdescription}}

\IEEEpeerreviewmaketitle





\IEEEpeerreviewmaketitle

%
\IEEEpeerreviewmaketitle

\section{Introduction}

\IEEEPARstart{H}{eating},  ventilation and air-conditioning (HVAC) system accounts for 40-50\% of  energy use in  buildings for providing occupant comfort~\cite{ku2015automatic, sun2013building}. Whereas a large group  are still  not actually  feeling comfortable staying inside~\cite{frontczak2012quantitative}.   This negative situation is substantially attributed to the insensible operation of HVAC systems,  such as settled on-off switch and fixed thermostat settings, etc. 
Fortunately,  the advances of information and communication technology (ICT) including smart sensing, data storing and processing technologies,  are likely to turn it around~\cite{ICTforBuildings}.
Advanced HVAC control is expected for energy-efficient buildings, i.e., reduce energy use (cost) while simultaneously enhancing human comfort~\cite{mossolly2009optimal, fasiuddin2011hvac}.

\vspace{-5mm}
\subsection{Literature}
The recent decades have seen  massive progress towards  improving building energy efficiency. 
Particularly, various modeling techniques have been established for optimizing HVAC control \cite{afroz2018modeling}. 
However,  the  \emph{system complexity} and \emph{uncertainties} are still two major challenges we are facing for realization. 
The HVAC system comprises various heat and mass transfer equipment such as the chiller, the heating/cooling coils and  the air-handling equipment, making the overall system dynamics  highly non-linear and non-convex~\cite{afroz2018modeling}.  
Besides, the HVAC control  requires to respond to the uncertain thermal disturbances caused by the weather and the occupants, etc.

To address the non-linearity,  the typical solution methods include  sequential quadratic programming (SQP) \cite{kelman2011bilinear}, mixed-integer programming (MIP)  \cite{xu2017pmv,xu2015supply} and meta-heuristic algorithms (e.g., genetic algorithm~\cite{garnier2015predictive} and particle swarm optimization~\cite{zou2010model}, etc.).  These works  have mostly focused on developing approximation or relaxation techniques to deal with the non-linear system dynamics. 
Besides, complete information or accurate predictions for dynamic thermal demand  are usually assumed for computation~(see \cite{kelman2011bilinear, xu2015supply} for examples).

However, the uncertainties caused by the weather and occupancy can not be underestimated  for practice. Both the weather and indoor occupancy fluctuate over the time, affecting the thermal demand to be responded by the HVAC operation.
In the literature, model predictive control (MPC) technique has been extensively discussed for HVAC control, especially deterministic MPC based on short-term predictions~\cite{afram2014theory}.  The basic idea of MPC is to rely on a model to predict the future system dynamic process and use the predictions to make a local optimal decision for current stage. 
However,  MPC is still being rarely used in buildings primarily  due to the two obstacles:  \emph{i)} the lack of appropriate model to be deployed in the MPC controller due to the system complexity~\cite{xu2017pmv}, and \emph{ii)} the difficulty to obtain accurate predictions for future thermal demand~\cite{ma2012predictive}. 
Moreover, deterministic MPC without considering the uncertainties of predictions has been found not work well  for HVAC control~\cite{oldewurtel2012use}. Aware of that,   stochastic MPC  \cite{oldewurtel2012use, oldewurtel2013stochastic}  and explicit MPC \cite{parisio2014control, klauvco2014explicit} have been suggested to  account for the (prediction) uncertainties.  They  minimize the average HVAC energy cost and provide thermal comfort with probability guarantees via  chance constraints.  These MPCs can  reduce  performance variance over  deterministic  MPCs,  but face the challenges of handling the chance constraints~\cite{kouvaritakis2015stochastic}, especially accounting for the on-line computation burden. 
To compensate the deficiency,  \cite{parisio2014control, klauvco2014explicit}  discussed explicit MPCs  for on-line HVAC control  with assumed linear  models.  


While  improving energy  efficiency of  HVAC systems has raised extensive attention, we are still on the way to achieve the target.
One one hand, a control method  that can handle the intrinsic non-linearity and uncertainties  is required.  On the other hand, the current practices have mostly used  static temperature ranges to indicate human thermal comfort (see \cite{kouvaritakis2015stochastic, oldewurtel2013stochastic}), which may not actually provide the thermal comfort warranted \cite{frontczak2012quantitative};  for the occupants' thermal sensation is determined by multiple parameters: indoor air temperature, mean radiant temperature, humidity, air velocity, metabolic activity and clothing insulation etc. \cite{fanger1970thermal}.  Therefore, it necessitates to integrate an elaborate thermal comfort model in the optimization.
Moreover, we can expect a more reasonable thermal comfort model can help save energy by avoiding over-cooling or over-heating~\cite{ku2015automatic, xu2017pmv}.  

\vspace{-3mm}
\subsection{Our Contributions}
To advance energy-efficient buildings,  this paper studies the optimal  HVAC control subject to uncertain thermal demand (i.e., weather and the occupants).   To enhance thermal comfort, we involve the elaborate PMV thermal comfort model~\cite{fanger1970thermal} in the optimization.  Particularly we deploy an Markov decision process (MDP) formulation~\cite{puterman2014markov, puterman1994markov} which is capable of accommodating the nonlinear system dynamics and the non-analytical PMV model for HVAC control but with the computational challenges to be addressed. 
To handle it, we propose a gradient-based learning ( GB-L) algorithm to  progressively learn the optimal stochastic  control policies off-line and store it for on-line execution.  We prove the method's  convergence to the optimal policies theoretically. Moreover, we demonstrate the performance (i.e., energy cost, thermal comfort and on-line computation) for HVAC control via simulations. The comparisons with  the existing MPC-based relaxation (MPC-R) method which is assumed with accurate future information and supposed to provide the near-optimal bounds, show though there exists some performance discount in energy cost reduction, the proposed method can enable efficient on-line implementation and high probability of thermal comfort  under uncertainties. 

The remainder is stuctured as:  Section II gives the MDP formulation, Section III  introduces the  GB-L method and  its convergence,  Section IV evaluates the  method  's performance for HVAC control via simulations, and Section V concludes this paper.

\vspace{-6mm}
\section{The Problem}
\vspace{-4mm}
\subsection{HVAC System Configuration}

An HVAC system mainly comprises  chiller and air handling units, with the latter integrated with  cooling/heating coils and driven fan.    
The HVAC system generally works in the way that 
the air handling units first  cool/heat and dehumidify the air to the set-points (temperature
and humidity) and then drive the supply air to the duct network connected to the rooms 
using the fans. Amid this process,  the cooling/heating coils rely on the chiller providing circulated  chilled water. Therefore, the chiller, cooling/heating coils and fan account for the major parts of  HVAC's energy consumption.   

\begin{figure}[h]
	\setlength{\abovecaptionskip}{-2pt}
	\setlength{\belowcaptionskip}{2pt}
	\centering
	\includegraphics[width=3.0in, height= 2.0in]{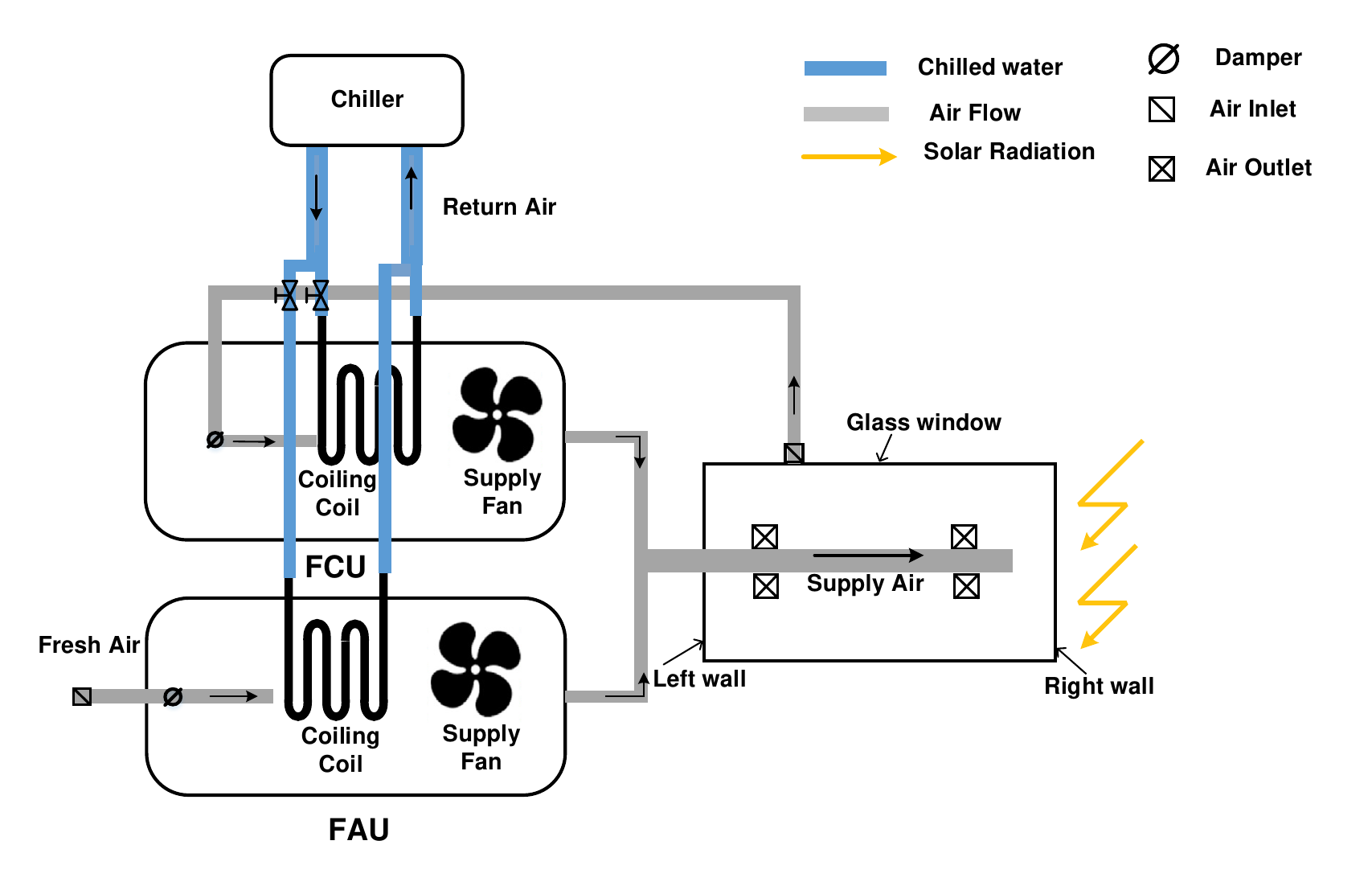}\\
	\vspace{-3mm}
	\caption{The configuration of the HVAC system for an office.}\label{HVAC system}
	\vspace{-3mm}
\end{figure}

In this paper, we study the type of  HVAC systems encompassing a fresh air unit (FAU)  and a fan coil unit (FCU) for handling the recirculated air and fresh air separately~\cite{sun2013building}.
As shown in Fig. \ref{HVAC system}, we focus on the HVAC control for a specific office for cooling
with the thermal demand mainly caused by the heat gain from the outside, indoor occupants  and electrical devices in use. 
To provide indoor comfort with least energy use, we study the optimal control inputs for  the HVAC system including supply air flow rates and the set-points of air handling units (i.e., FAU and FCU).
We note  there exist other types of HVAC systems that have integrated the FAU and FCU into one  unit~\cite{radhakrishnan2017learning}. 
This paper just uses the type of  HVAC system as an example to build a general framework,   and the formulation and  method  suppose to be extended accordingly. 

Since the building thermal dynamics is a slow process and the arrival/ departure of occupants are spontaneous, we discuss the problem in a discrete-time setting with a sampling and control interval of $\Delta t$ = 30 mins over a daily circle (i.e., $T=48$ stages).

\subsection{MDP Formulation}

A standard MDP is composed of system states, control, performance  function and  system dynamics. In the following, we give their specific definitions for HVAC control. 

\subsubsection{System state}
The system state is the input for computing HVAC control at each stage.  Obviously, it should characterize the cooling demand  determined  by the present indoor condition ($T^{\text{a}}_t, H^{\text{a}}_t$), the weather ($T^{\text{o}}_t, H^{\text{o}}_t$)  and the occupancy ($N^{\text{a}}_t$): $$S_t=[T^{\text o}_t, H^{\text o}_t,  T^{\text a}_t,  H^{\text a}_t,  N^{\text a}_t]^{\top}$$

\subsubsection{Control variables}
Since the   air flow rates  and  the set-points of FAU and FCU  are to be decided,  we  have  the control: 
$$A_t=[G^{\text{fau}}_t, T^{\text{fau}}_t, G^{\text{fcu}}_t, T^{\textrm{fcu}}_t]^{\top}$$

\subsubsection{System dynamics}
We use the  gray-box model  deduced from the energy and mass conservation equations to capture the indoor  thermal and humidity dynamics~\cite{wu2016optimal, sun2013building} and Markov chains to model the weather and indoor occupancy patterns.

\emph{Temperature:} ~ The indoor temperature variation  is  the interplay of   HVAC control,  thermal disturbance (i.e., the weather and occupancy),  and the building thermal inertia, which can be described by the equations  \cite{wu2016optimal, sun2013building}:

\vspace{-3mm}
{\small{
		\begin{equation} \label{indoor temperature}
		\begin{split}
		C_{\text{p}} &m^{\text a} (T^{ \text a}_{t+1}-T^{\text a}_t)= N^{\text a}_t  (Q^{\text o} + Q^{\text d})\Delta_t+  h_{\text{gs}} A_{\text{gs}}(T^{\text o}_t-T^{\text a}_t)\Delta_t \\
		&+ h_{\text w}A_{\text{wl}}(T^{\text{wl}}_t-T^{\text a}_t) \Delta_t + h_{\text w} A_{\text{wr}}(T^{\text{wr}}_t-T^{\text a}_t) \Delta_t  \\
		&+ G^{\text{fau}}_t (T^{\text{fau}}_t-T^{\text a}_t)\Delta_t +G^{\text{fcu}}_t (T^{\text{fcu}}_t-T^{\text a}_t)\Delta_t\\
		\end{split}
		\end{equation}}}
where the  first  term  captures  the heat generated by the occupants and electrical devices  (e.g., laptops, monitors and desktops etc.).
The  next  three terms calculate  the heat gains from the glass window and  walls, and the last two  model the cooling power provided by the FAU and FCU. 

As \eqref{indoor temperature} shows,  the inside and outside interact through the walls. We have  the wall thermal dynamics:

\vspace{-3mm}
{\small{
		\begin{equation}
		\begin{split}
		&C_{\text w} m^{\text{wl}}(T^{\text{wl}}_{t+1}-T^{\text{wl}}_t)=h_{\text{w}} A_{\text{wl}}(T^{\text{a}}_t- T^{\text{wl}}_t)\Delta_t\\
		&C_{\text w} m^{\text{wr}}(T^{\text{wr}}_{t+1}-T^{\text{wr}}_t)=h_{\text w} A_{\text{wr}}(T^{\text a}_t- T^{\text{wr}}_t)\Delta_t+ \alpha_{\text w} A_{\text{wr}} Q^{\text w}_t\Delta_t 
		\end{split}
		\end{equation}}}

\vspace{-0.1cm}

\emph{Humidity:}~  According to~\cite{sun2013building},  we have the indoor humidity dynamics: 

\vspace{-3mm}
{\small{
		\begin{equation} \label{humidity}
		\begin{split}
		m^{\text a}(H^{\text a}_{t+1}-H^{\text a}_t)=& N^{\text a}_t  H^{\text g}\Delta_t+ G^{\text{fau}}_t  (H^{\text{fau}}_t-H^{\text a}_t)\Delta_t\\
		&+G^{\text{fcu}}_t (H^{\text{fcu}}_t-H^{\text a}_t)\Delta_t
		\end{split}
		\end{equation}}}
where the first term calculates the humidity exhaled by the occupants,  and the other two  model the dehumidifying of FAU and FCU for the circulated air.  The dehumidifying intensity of FAU and HCU  are determined by their saturation set-points,   and we have  
$H^{\textrm{fau}}_t=\min(H^{\text o}_t, H^{\text{fau}, \text{sat}})$ and $H^{\textrm{fcu}}_t=\min(H^{\text a}_t, H^{\text{fcu}, \text{sat}})$.

\emph{Occupancy:}~ We use Markov chain to capture the occupancy patterns~\cite{jia2017privacy, shen2017leveraging}:

\vspace{-3mm}
{\small{
		\begin{equation}\label{indoor occupancy}
		\begin{split}
		\textrm{Pr}&(N^{\text a}_{t+1}=j|N^{\text a}_t=i)=\bm{P}_t^{\text{N}}[i,j], ~\forall i, j, \in \{1, \cdots, L_{\text N}\}.
		\end{split}
		\end{equation}}}
where $\bm{p}^{\text{N}}_t \in \mathbb{R}^{L_{\text N}}$ denotes  the transition probability matrix of occupancy at time $t$. The integers  $i, j$ represents the  occupancy levels within the $L_{\text N}$ segments.

\emph{Weather condition:} ~ Similarly, we  uses Markov chains  to capture  the outdoor  temperature and humidity dynamics: 

\vspace{-3mm}
{\small{
		\begin{equation} \label{weather dynamics}
		\begin{split}
		\textrm{Pr}(T_{t+1}^{\text{o}}&=j | T^{\text o}_{t+1}=i)=\bm{P}^{\text{T}}_t[i,j], ~\forall i, j \in \{1, \cdots, L_{\text T}\}.\\
		\textrm{Pr}(H_{t+1}^{\text o}&=j | H^{\text o}_{t+1}=i)=\bm{P}^{\text H}_t[i, j], ~\forall i, j \in \{1, \cdots, L_{\text H}\}.\\
		\end{split}
		\end{equation}}}
where $\bm{P}^{\text T}_t$ and  $\bm{P}^{\text H }_t$ represent the transition probability matrices.   In particular, we equally divide   the  outdoor temperature and humidity ranges into  $L_{\text T}$ and $L_{\text H}$ segments.

\subsubsection{Objective function} Considering the energy use is hard to inspect in practice,   we select the electricity bill of HVAC system as the objective: 

\vspace{-3mm}
{\small
	\begin{equation} \label{objective function}
	\begin{split}
	J\!=\!\mathbb{E}\Big[\sum_{t=0}^{T-1} c_t  \big\{ \eta (C^{\textrm{fcu}}_t \!+\!C^{\textrm{fau}}_t) \!+\!F^{\textrm{fcu}}_t\!+\!F^{\textrm{fau}}_t\big\}\Delta_t \Big],
	\end{split}
	\end{equation}}
where  the  cooling power {\small $C^{\textrm{fcu}}_t, C^{\textrm{fau}}_t$}  and  the fan power   {\small $F^{\text{fcu}}_t, F^{\text{fau}}_t$} for  the FAU and FCU are calculated as  \cite{sun2013building}

\vspace{-3mm}
{\small
	\begin{subequations} 
		\begin{alignat}{4}
		&C^{\textrm{fcu}}_t\!=\!C_{\text p} G^{\textrm{fcu}}_t (T^{\text a}_t-T^{\textrm{fcu}}_t)+C_{\text p} G^{\textrm{fcu}}_t \big[  H^{\text a}_t (2500+1.84T^{\text a}_t) \notag\\
		&\quad \quad \quad- H^{\text{fcu}}_t (2500+1.84T^{\text{fcu}}_t)\big]\\
		&C^{\textrm{fau}}_t\!=\! C_{\text p} G^{\textrm{fau}}_t (T^{\text o}_t-T^{\textrm{fau}}_t )\!+\!C_{\text p} G^{\textrm{fau}}_t \big[H^{\text o}_t (2500+1.84T^{\text o}_t) \notag\\
		&\quad \quad \quad-(2500+1.84T^{\textrm{fau}}_t)\big]\\
		&\label{eq:9c} F^{\text{fcu}}_t =F^{\text{fcu}, \textrm{r}} \big( { G^{\text{fcu}}_t }/{G^{\text{fcu}, \text{r}}} \big)^3\\
		&\label{eq:9d} F^{\text{fau}}_t=F^{\text{fau}, \textrm{r}}\big( {G^{\text{fau}}_t}/{G^{\text{fau}, \text{r}}}\big)^3
		\end{alignat}
\end{subequations}}
\vspace{-0.3cm}

\subsubsection{PMV Model} To enhance thermal comfort,  this  paper capitalizes on  the elaborate PMV model   to capture the occupants' satisfaction~\cite{fanger1970thermal}.  The PMV model is non-analytical and implicitly characterized by a number of equations. For brevity,  we denote the PMV model as 

\vspace{-3mm}
{\small
	\begin{equation} \label{PMV model}
	\textrm{pmv}_t=\text{PMV}(M, W,  T^{\text a}_t,  H^{\text a}_t, t^{\text r}_t, v^{\text a}_t, I_{\text{cl}})
	\end{equation}}
where the inputs  include: metabolic rate $M$ (\si{\watt \per\milli^2}), mechanic work intensity $W$  (\si{\watt \per\milli^2}), indoor air temperature $T^{\text a}_t$ (\si{\degreeCelsius}),  relative humidity $H^{\text a}_t$ (\si{\percent}), 
mean radiation temperature $t^{\text r}_t$ (\si{\degreeCelsius}),  indoor air velocity $v^{\text a}_t$ (\si{\milli^2}), and  clothing insulation $I_{\text{cl}}$ (\si{\milli^2 \kelvin \per \watt}). 
The PMV model  establishes   the  mapping from  indoor condition   to the occupants'  average satisfaction within the range $[-3, 3]$. Particularly, the numbers $-3$, $0$,  $3$ indicate  too cold, ideal, and  too hot, respectively.

\normalsize
\subsubsection{Constraints} The operation of the HVAC  system should comply with  the physical limits of  dampers within the FAU and FCU \eqref{eq:11a}, as well as the chiller capacity that determines the temperature set-points  \eqref{eq:12a}. 

\vspace{-3mm}
\small
\begin{subequations} \label{air flow rate constraints}
	\begin{alignat}{4}
	\label{eq:11a} &\underline{G}^{\textrm{fau}}\leq G^{\textrm{fau}}_t  \leq \overline{G}^{\textrm{fau}},~
	\underline{G}^{\textrm{fcu}}\leq G^{\textrm{fcu}}_t  \leq \overline{G}^{\textrm{fcu}}. \\
	\label{eq:12a} &	\underline{T}^{\textrm{fau}}\leq T^{\textrm{fau}}_t  \leq \overline{T}^{\textrm{fau}},~
	\underline{T}^{\textrm{fcu}}\leq T^{\textrm{fcu}}_t  \leq \overline{T}^{\textrm{fcu}}.
	\end{alignat}
\end{subequations}
\normalsize

We use $\underline{\textrm{pmv}}$ and $\overline{\textrm{pmv}}$ to capture the human comfort requirement  characterized by the PMV value, i.e., 
\begin{equation} \label{thermal comfort}
\begin{split}
\underline{\textrm{pmv}} \leq &\textrm{pmv}_t \leq \overline{\textrm{pmv}}
\end{split}
\end{equation}
where we usually have $\underline{\textrm{pmv}} =- 0.5$ and $\overline{\textrm{pmv}}=0.5$.

\subsubsection{Optimization problem}
Overall, the optimal control of the HVAC system to optimize  energy cost while respecting occupant comfort can be described as 

\vspace{-3mm}
\small
\begin{eqnarray} \label{Optimization Problem}
\begin{split}
&\min_{\pi}~ J(S_0, \pi)\!=\!\mathbb{E}^{\pi}\Big[\sum_{t=0}^{T-1} c_t  \big\{ \eta (C^{\textrm{fcu}}_t +C^{\textrm{fau}}_t) +F^{\text{fcu}}_t+F^{\text{fau}}_t\big\}\Delta_t \Big]\\
& s.t.~  \quad\textrm{{System dynamics:}} (\ref{indoor temperature})-(\ref{weather dynamics}),  ~\textrm{{Operation limits:} (\ref{air flow rate constraints}}), \\
&\quad \quad  \quad  \textrm{{Thermal comfort:}}~ (\ref{thermal comfort}),~\forall t\in \{0, 1, \cdots, T-1\}.
\end{split}
\end{eqnarray}
\normalsize
where  $\pi=(\pi_0, \pi_1, \cdots,  \pi_{T-1})$ denotes  the control policy over the optimization horizon. $S_0$ is the initial system state. At each time $t$, the control  $\pi_t$: $\mathcal{S}_t \rightarrow \mathcal{A}_t$ establishes a mapping from the sate space  $\mathcal{S}_t$ to the action space $\mathcal{A}_t$.  $\mathbb{E}^{\pi}$ denotes the expectation under policy $\pi$.

We have formulated the optimal control of HVAC system as a standard MDP \eqref{Optimization Problem}.
However, searing  for  an optimal policy is nontrivial and remains to be addressed.  Dynamic programming  (DP)  for problem \eqref{Optimization Problem}  requires to transverse  Q-factors for the Cartesian product of state and action space backward or forward to identify the optimal deterministic policy~\cite{puterman1994markov}. This is computationally impractical due to \emph{i}) the multiple uncertainties  resulting in complicated state transitions,  and \emph{ii)} the large state and action space posing  intensive computation.





%

\section{Gradient-based Learning}
To handle \eqref{Optimization Problem},  we propose a  gradient-based learning (GB-L) algorithm to 
progressively learn  a stochastic policy off-line and store it for on-line implementation.  To account for the multiple uncertainties,  the Monte Carlo (MC)  technique  \cite{jia2012simulation}  is employed to simulate system dynamics and estimate the performance gradients of policies. 
Particularly,  for the HVAC control,  we establish two specific strategies for reducing computation by exploring the problem features.  

\vspace{-0.5cm}
\subsection{Notations}
We use  the lower cases $\bm{s}_t$ and $\bm{a}_t, \bm{b}_t, \bm{c}_t$ to represent state and action instances at time $t$, respectively. We use the integer sets  $\mathcal{S}_t \triangleq  \{1, 2, \cdots, |\mathcal{S}_t| \}$  and  $  \mathcal{A}_t \triangleq \{1, 2, \cdots, |\mathcal{A}_t| \}$ to denote the state  and action space, where the operator $\vert \cdot \vert$ defines the cardinality.  $\bm{\theta}=(\theta_0, \theta_1, \cdots, \theta_{T-1})^T$ denotes a stochastic policy, where  $\theta_t \in \mathbb{R}^{|\mathcal{S}_t| \times |\mathcal{A}_t|}$ establishes  the mapping from the state space $\mathcal{S}_t$  to the action space $\mathcal{A}_t$ with
$\theta_t (\bm{s}_t, \bm{a}_t) \in [0, 1]$ denoting the probability to take action $\bm{a}_t$  at  state $\bm{s}_t$. 
The  lower cases  $p_t(\bm{s}_{t+1}|\bm{s}_t, \bm{a}_t)$ and $p^{\bm{\theta}}_t (\bm{s}_{t+1}|\bm{s}_t)$   indicate  the state transition probability when taking  action $\bm{a}_t$ or  policy $\bm{\theta}$, and  $\bm{P}^{\theta}_t=[p_t^{\bm{\theta}}(\bm{s}_{t+1}|\bm{s}_t)] \in \mathbb{R}^{ |\mathcal{S}_t| \times  |\mathcal{S}_{t+1} | }$ is  the transition probability matrix.  The superscript $k$  denotes the iteration. 
\vspace{-0.5cm}
\subsection{GB-L}
To handle the various constraints of problem \eqref{Optimization Problem}, we redefine an augmented stage-cost function:
{\small{
		\begin{equation} \label{stage-cost}
		\begin{split}
		r_t(\bm{s}_t, \bm{a}_t)=&c_t  \big[ \eta (C^{\textrm{fcu}}_t +C^{\textrm{fau}}_t) +F^{\textrm{fcu}}_t+F^{\textrm{fau}}_t\big] \Delta_t +I_t(\mathcal{X}_t)
		\end{split}
		\end{equation}}}
where the first part is the energy cost  and the second part quantifies the penalty of constraint violations. Particularly, we use $\mathcal{X}_t$ to indicate the set of constraints at time $t$, comprising  \eqref{indoor temperature}-\eqref{weather dynamics}, \eqref{air flow rate constraints} and \eqref{thermal comfort}.
We have the indicator function  $I_t(\cdot)$  that $I(A)=1$ with the condition $A$ true, otherwise $I(A)=0$.


Our method  is inspired  by the performance difference equation  \cite{Zhao2010} that for any two stochastic policies $\bm{\sigma}$ and $\bm{\mu}$, the induced  performance difference can be quantified by 

\vspace{-3mm}
{\small
	\begin{equation} \label{performance difference}
	\begin{split}
	&J(\bm{\mu}; S_0 )\!-\!J(\bm{\sigma}; S_0 )\!=\!\!\sum_{t=0}^{T-1} \pi^{\bm{\mu}}_t \Big[ (\bm{r}^{\bm{\mu}}_t\!-\!\bm{r}^{\bm{\sigma}}_t)\!+\!(\bm{P}^{\bm{\mu}}_t\!-\!\bm{P}^{\bm{\sigma}}_t)V^{\bm{\sigma}}_{t+1}\Big]
	\end{split}
	\end{equation}}
where we have 
\begin{equation*}
\begin{split}
& \bm{\pi}^{\bm{\theta}}_t=(\pi^{\bm{\theta}}_t(1), \pi^{\bm{\theta}}_t(2), \cdots, \pi^{\bm{\theta}}_t(|\mathcal{S}_t|))^\top,  \\
& \bm{r}_t^{\bm{\theta}}=(r^{\bm{\theta}}_t(1), r^{\bm{\theta}}_t(2), \cdots, r^{\bm{\theta}}_t(|\mathcal{S}_t|))^\top,  \\
& V^{\bm{\theta}}_{t+1}\!=\!(V^{\bm{\theta}}_{t+1}(1), \!\cdots,\! V^{\bm{\theta}}_{t+1}(|\mathcal{S}_{t+1}|))^\top,
\end{split}
\end{equation*}
denote the state distribution, stage-cost and performance potential for policy $\theta  \in \{\bm{\mu}, \bm{\sigma}\}$. 
In particular, the performance potential is defined as 
\begin{equation}
\begin{split}
V^{\bm{\theta}}_{t+1}(\bm{s}_{t+1})=\mathbb{E}^{\bm{\theta}}  \big[\sum_{\tau=t+1}^{T-1}  r_{\tau} (s_\tau, a_\tau) \big], \textrm{with}~ a_{\tau}=\bm{\theta}(s_{\tau}).
\end{split}
\end{equation} 

From the perspective of perturbation analysis (PA),  the stochastic policy 
$\bm{\sigma}$ and $\bm{\mu}$ can be viewed as the base and perturbed policy. 
The intuitive interpretation of \eqref{performance difference} is that the performance of the perturbed policy $\bm{\mu}$ can be constructed by the performance potentials  $V^{\bm{\sigma}}_{t+1}(\bm{s}_{t+1})$  of   base policy $\bm{\sigma}$. However, it requires  the explicit formula of  state distribution $\bm{\pi}^{\bm{\mu}}$ under the perturbed policy $\bm{\mu}$, which is generally unknown \emph{a priori}, making it impractical to use (\ref{performance difference}) for policy update.

To handle it, we derive a differential formula 
by defining a structured  random policy $\bm{\delta}$ which adopts policy $\bm{\sigma}$ with  probability $\delta$ and policy $\bm{\mu}$ 
with  probability $1\!\!-\!\!\delta$.  For policy $\bm{\delta}$, we can easily identify  the state  transition probability  matrix $\bm{P}^{\bm{\delta}}_t=\bm{P}^{\bm{\sigma}}_t+\delta \Delta \bm{P}_t$ with  $\Delta \bm{P}_t\!=\!\bm{P}^{\bm{\mu}}_t\!-\! \bm{P}^{\bm{\sigma}}_t$,  and the stage-cost vectors $\bm{r}^{\bm{\delta}}_t\!=\!\bm{r}^{\bm{\sigma}}_t+\delta \Delta \bm{r}_t$ with $\Delta \bm{r}_t\!=\!\bm{r}^{\bm{\mu}}_t-\bm{r}^{\bm{\sigma}}_t$.   In this regard, the equation \eqref{performance difference} can be translated into

\vspace{-0.3cm}
{\small{
		\begin{equation} \label{performance difference2}
		\begin{split}
		J(&\bm{\delta}; S_0 )-J({\bm{\sigma}}; S_0 )=\sum_{t=0}^{T-1} \pi^{\bm{\delta}}_t \Big[ \Delta \bm{r}_t+\Delta \bm{P}_t \bm{V}^{\bm{\sigma}}_{t+1}\Big]
		\end{split}
		\end{equation}}}

\vspace{-4mm}
By assuming $\delta\rightarrow0$, we  have  the  performance differential equation:

\vspace{-3mm}
{\small{
		\begin{equation} \label{differential equation}
		\frac{d J(\bm{\delta}; S_0)}{d \delta}=\lim\limits_{\delta\rightarrow 0} \sum_{t=0}^{T-1} \pi^{\bm{\delta}}_t \Big[ \Delta \bm{r}_t+\Delta \bm{P}_t \bm{V}^{\bm{\sigma}}_{t+1}\Big]
		\end{equation}}}
\vspace{-0.2cm}
Equivalently,  we have 
\small
\begin{equation} \label{difference equation}
\begin{split}
&\frac{\partial J(\bm{\sigma}; S_0)}{\partial \sigma_t(\bm{s}_t, \bm{a}_t)}\!=\!\pi_t^{\bm{\sigma}}(\bm{s}_t) \Big[\frac{\partial r^{\bm{\sigma}}_t(\bm{s}_t)}{\partial \sigma_t(\bm{s}_t, \bm{a}_t)}+\sum_{\mathclap{{\bm{s}_{t+1}\in  \mathcal{S}_{t+1}}}}\frac{\partial p^{\bm{\sigma}}_t (\bm{s}_{t+1}|\bm{s}_t)}{\partial \sigma_t (\bm{s}_t, \bm{a}_t)} \bm{V}^{\bm{\sigma}}_{t+1}(\bm{s}_{t+1})\Big]
\end{split}
\end{equation}
\normalsize

{\small{By substituting  $r^{\bm{\sigma}}_t(\bm{s}_t)\!\!\!=\!\!\!\sum_{\bm{a}_t=1}^{|\mathcal{A}_t|}p_t^{\sigma}(\bm{a}_t|\bm{s}_t) r_t(\bm{s}_t, \!\bm{a}_t)$, $p^{\bm{\sigma}}_t (\bm{s}_{t+1}|\bm{s}_t)$  $\!\!=\!\!\sum_{\bm{a}_t=1}^{ |\mathcal{A}_t |}\! p_t^{\bm{\sigma}}(\bm{a}_t|\bm{s}_t) p(\bm{s}_{t+1}|\bm{s}_t, \bm{a}_t )$ into \eqref{difference equation}, we have}}

{\small{
		\begin{equation} \label{(24)}
		\begin{split}
		&\frac{\partial J(\bm{\sigma}; S_0)}{\partial \sigma_t(\bm{s}_t, \bm{a}_t)}=\pi_t^{\bm{\sigma}}(\bm{s}_t)\Big[\sum_{\bm{b}_t=1}^{|\mathcal{A}_t|} \frac{\partial p_t^{\bm{\sigma}} (\bm{b}_t|\bm{s}_t)}{\partial \sigma_t (\bm{s}_t, \bm{a}_t)} r_t(\bm{s}_t, \bm{b}_t)\\
		&+\sum_{{\bm{s}_{t+1}\in  \mathcal{S}_{t+1}}}\sum_{\bm{b}_t=1}^{|\mathcal{A}_t|} \frac{\partial p_t^{\bm{\sigma}}(\bm{b}_t|\bm{s}_t)}{\partial \sigma_t(\bm{s}_t, \bm{a}_t)} p_t(\bm{s}_{t+1}|\bm{s}_t, \bm{b}_t) \bm{V}^{\bm{\sigma}}_{t+1}(\bm{s}_{t+1})\Big]\\
		\end{split}
		\end{equation}
}}
$$ =\pi^{\bm{\sigma}}_t (\bm{s}_t) \Big [\sum_{\bm{b}_t=1}^{|\mathcal{A}_t|} \frac{\partial p_t^{\bm{\sigma}}(\bm{b}_t|\bm{s}_t)}{\partial \sigma_t(\bm{s}_t, \bm{a}_t)}\big(r_t(\bm{s}_t, \bm{b}_t)+V^{\bm{\sigma}}_{t}(\bm{s}_t, \bm{b}_t)\big) \Big]$$
where   $V^{\bm{\sigma}}_{t}(\bm{s}_t, \bm{b}_t)=\sum_{{\bm{s}_{t+1}\in  \mathcal{S}_{t+1}}}p_t(\bm{s}_{t+1}|\bm{s}_t, \bm{b}_t) V^{\bm{\sigma}}_{t+1}(\bm{s}_{t+1})$.

As $p_t^{\bm{\sigma}}(\bm{b}_t|\bm{s}_t)=\frac{\sigma_t(\bm{s}_t, \bm{b}_t)}{\sum_{\bm{c}_t=1}^{|\mathcal{A}_t|} \sigma_t (\bm{s}_t, \bm{c}_t)}$, we have

\small
\begin{equation}  \label{(22)}
\frac{\partial p_t^{\bm{\sigma}}(\bm{b}_t|\bm{s}_t)}{\partial \sigma_t(\bm{s}_t, \bm{a}_t)}\!=\!\left\{
\begin{aligned}
&\frac{\sum_{\bm{c}_t=1}^{|\mathcal{A}_t|} \sigma_t(\bm{s}_t, \bm{c}_t)\!-\!\sigma_t(\bm{s}_t, \bm{a}_t)}{[\sum_{\bm{c}_t=1}^{|\mathcal{A}_t|} \sigma_t(\bm{s}_t, \bm{c}_t)]^2}, \quad \quad  &\bm{b}_t=\bm{a}_t  \\
&\frac{-\sigma(\bm{s}_t, \bm{b}_t)}{[\sum_{\bm{c}_t=1}^{|\mathcal{A}_t|} \sigma_t(\bm{s}_t, \bm{c}_t)]^2},  \quad \quad  &\bm{b}_t\neq \bm{a}_t\\
\end{aligned}
\right.
\end{equation}
\normalsize

The equation (\ref{(24)}) can be  interpreted as the performance  gradients  of  policy $\bm{\sigma}$,  and thus we can establish the standard policy update procedure: 
{\small{
		\begin{equation} \label{policy iteration}
		\begin{split}
		\bm{\sigma}^{k+1}=\bm{\sigma}^k-\bm{\gamma}^k \cdot \nabla_{\bm{\sigma}^k} J(\bm{\sigma}^k; S_0)
		\end{split}
		\end{equation}}}
where we have {\small$\nabla_{\bm{\sigma}^k}J(\bm{\sigma}^k; S_0)=\big[\frac{\partial J(\bm{\sigma}^k; S_0)}{\partial {\sigma}^k_t(\bm{s}_t, \bm{a}_t)}\big]$. $\bm{\gamma}^k=[\gamma_t^k(\bm{s}_t, \bm{a}_t)]$}  denotes the step-size at iteration $k$.

For the policy update formula \eqref{policy iteration},  we note  two  problems to be addressed:  
\emph{i)} computing the performance gradients $\nabla_{\bm{\sigma}^k}J(\bm{\sigma}^k; S_0)$, and
\emph{ii)} determining the step-size $\bm{\gamma}^k$. 
As there exists randomness both for the weather and occupancy,  it's impractical to analytically estimate the performance gradients  under expectation.   To overcome this difficulty,  the MC method \cite{jia2012simulation} is used to simulate the system dynamics and  estimate 
the performance gradients $\nabla_{\bm{\sigma}}J(\bm{\sigma}^k; S_0)$ for any given policy  $\bm{\sigma}$. 
The  main procedures are shown  in \textbf{Algorithm} \ref{GradientsEstimation}. 
We use $\Omega_t(\cdot)$ to denote the sets of states or actions visited at time $t$ and $\mathcal{I}_t(\cdot)$ record the sample indices accordingly.
The algorithm  consists of  two  main steps:
\emph{i)} generate a number of sample paths by  executing the policy $\bm{\sigma}$ via MC simulations, and 
\emph{ii)} estimate the state distribution $\pi^{\bm{\sigma}}_t(\bm{s}_t)$, stage-cost $r_t( \bm{s}_t, \bm{b}_t )$ and performance potentials $V^{\bm{\sigma}}_t(\bm{s}_t, \bm{b}_t )$.    
The complete implementation of GB-L is shown in \textbf{Algorithm} \ref{GB-L} which comprises  the two main steps: \emph{i)} estimate the performance gradients (\textbf{Step} 3), and \emph{ii)} update the policy (\textbf{Step} 4). 
We select $\lVert \nabla_{\bm{\sigma}^k} J(\bm{\sigma}^k; S_0)\rVert_2\leq \epsilon$ ($\epsilon$  is positive threshold) as the stopping criterion.  Note that the learning process can be carried out off-line and  store the obtained policy for on-line implementation. In such setting, we can expect efficient on-line execution as it only requires to identify the action at each stage  based on the system states. 
For the step-size $\bm{\gamma}^k$, which is  closely related to  the method's converge,   we have the main results in \emph{\textbf{Theorem}} 1. 

\begin{theorem} \label{theorem1} For any given initial policy $\bm{\sigma}^0$, the  GB-L method
	can converge to an optimal policy of problem (\ref{Optimization Problem}) with the  selected stepsize $\gamma_t^k(\bm{s}_t, \bm{a}_t)=\frac{\sigma_t^k(\bm{s}_t, \bm{a}_t)}{\sum_{\bm{c}_t=1}^{|\mathcal{A}_t|} \sigma_t^k(\bm{s}_t, \bm{c}_t)}$ ($\forall \bm{s}_t \in \mathcal{S}_t$, $\bm{a}_t \in \mathcal{A}_t $) and  the  performance gradients  $\nabla_{\bm{\sigma}^k} J(\bm{\sigma}^k; S_0)$ estimated accurately enough. 
\end{theorem}
The proof refers to \textbf{Appendix} A. 

\begin{remark}
	As \textbf{Theorem} 1 illustrates,  the method's convergence  depends on the estimation accuracy  of the performance gradients.  Generally, any accurate estimation can be approached  by  increasing the number of sample paths.  However, we want to explain that the practical implementation only requires an estimation that preserve the (performance) order of the (action) candidates.
	
	An example: For any  two action $a$ and $b$, we only require an estimation that $\tilde{J}(a)\leq \tilde{J}(b)$ if $J(a)\leq J(b)$   where $J(\cdot)$ and $\tilde{J}(\cdot)$ denote the real and estimated value.   
	
\end{remark}

\begin{remark}
	The method's convergence  does not depend on the selection of  initial  policy $\bm{\sigma}^0$, which can be identified from the proof of \textbf{Theorem} \ref{theorem1}.  However,  a better initial policy  is expected  to yield faster convergence  and less iterations. 
\end{remark}


\begin{algorithm}[h] 
	\caption{Estimate the Performance Gradients via MC Simulation} \label{GradientsEstimation}
	\begin{algorithmic}[1]
		\State \textbf{Input}: a given stochastic policy $\bm{\sigma}$. 
		\State  Generate $\vert \mathcal{W} \vert $ 
		sample paths by executing policy $\bm{\sigma}$ and label the sample paths
		{\small 
			\begin{equation*}
		     \big\{ \bm{s}^\omega_0, \bm{a}^\omega_0, r^\omega_0, \bm{s}^\omega_1, \bm{a}^\omega_1,  r^{\omega}_1\cdots, \bm{s}^\omega_{T-1},  \bm{a}^\omega_{T-1}, r_{T-1}^{\omega} \big\}, \forall \omega \in \mathcal{W}.
	     \end{equation*}}
		\State \textbf{For} $t \in \{0, 1, \cdots, T-1\}$
		\State \quad Record the occurrence of  states in the sample paths:
		\small 
		\begin{equation*}
		\begin{split}
		&\Omega_t(\bm{s}_t)=\{s^{\omega}_t|~ \omega \in \mathcal{W}\}.\\
	    &\Omega_t( \bm{a}_t| \bm{s}_t)=\{ \bm{a}^{\omega}_t|~ \bm{s}^{\omega}_t=\bm{s}_t, \omega \in  \mathcal{W}\}, ~\forall \bm{s}_t \in \Omega_t(\bm{s}_t).\\
		& \mathcal{I}_t( \bm{s}_t)=\{\omega |~ \omega \in  \mathcal{W} \},  s^\omega_t=\bm{s}_t\}, ~ \forall \bm{s}_t \in \Omega_t(\bm{s}_t). \\
		  	& \mathcal{I}_t( \bm{s}_t, \bm{a}_t)=\{\omega |~ \omega \in \mathcal{W},  s^\omega_t=\bm{s}_t, a^{\omega}_t=\bm{a}_t\}, \\
		  &\quad\quad \quad  \quad \quad ~\forall  \bm{s}_t \in \Omega_t(\bm{s}_t),~\bm{a}_t \in \Omega_t( \bm{a}_t| \bm{s}_t).
		\end{split}
		\end{equation*}
		\normalsize
		\State Estimate  $\pi^{\bm{\sigma}}_t(\bm{s}_t) $, $r_t(\bm{s}_t, \bm{a}_t) $ and $V_t^{\bm{\sigma}}(\bm{s}_t, \bm{a}_t)$ according to
		{\small 
		\begin{equation*}
		\begin{split}
		&\pi^{\bm{\sigma}}_t(\bm{s}_t) \approx {|\mathcal{I}(\bm{s}_t)|}/{\vert\mathcal{W}\vert} , ~\forall \bm{s}_t \in \Omega_t(\bm{s}_t).\\
		& r_t(\bm{s}_t, \bm{a}_t) \approx \frac{1}{\vert\mathcal{I}(\bm{s}_t, \bm{a}_t)\vert} \sum_{\bm{\omega} \in \mathcal{I}(\bm{s}_t, \bm{a}_t)}r^{\bm{\omega}}_t, \\
		&\quad \quad ~\forall \bm{s}_t \in \Omega_t(\bm{s}_t), \bm{a}_t \in\Omega_t(\bm{a}_t|\bm{s}_t). \\
		&V_t^{\bm{\sigma}}(\bm{s}_t, \bm{a}_t)\approx  \frac{1}{\vert\mathcal{I}(\bm{s}_t, \bm{a}_t)\vert} \sum_{\bm{\omega} \in \mathcal{I}(\bm{s}_t, \bm{a}_t)}\sum_{\tau=t+1}^{T-1}r^{\bm{\omega}}_\tau,\\
		&\quad \quad  ~\forall \bm{s}_t \in \Omega_t(\bm{s}_t), \bm{a}_t \in\Omega_t(\bm{a}_t|\bm{s}_t). \\
		\end{split}
		\end{equation*}}
		\normalsize
		\State Compute $\nabla_{\bm{\sigma}}J(\bm{\sigma}; S_0)$ based on \eqref{(24)} and \eqref{(22)}.
		\State \textbf{EndFor}
	\end{algorithmic}
\end{algorithm}
	\vspace{-3mm}
{\small 
\begin{algorithm}[h] 

	\caption{Gradient-based Learning (GB-L)} \label{GB-L}
	\begin{algorithmic}[1]
		\State \textbf{Initialization:} $k \rightarrow 0$, $\bm{\sigma}^0$.
		\State \textbf{Iteration:}
		\State Estimate the gradients $\nabla_{\bm{\sigma}^k} J(\bm{\sigma}^k; S_0)$ using \textbf{Algorithm} \ref{GradientsEstimation}.
		\State Policy Update: 
		\begin{equation}
		\bm{\sigma}^{k+1}=\bm{\sigma}^k-\bm{\gamma^k} \cdot \nabla_{\bm{\sigma}^k} J(\bm{\sigma}^k; S_0)
		\end{equation}
		\State Stop if the \textbf{stopping criterion} \eqref{stopping} is reached, otherwise  $k = k+1$ go to \textbf{Step} 3.
	\end{algorithmic}
\end{algorithm} }

\vspace{-5mm}
\subsection{Computation Reduction}
We note the main computation burden of GL-B lies in estimating the performance gradients of policies (\textbf{Step} 3 of \textbf{Algorithm} \ref{GB-L}). According to \eqref{(24)},  the performance potentials of  the state-action pairs are required to build the performance gradients. 
For HVAC control, both the state and control encompass concatenated variables, causing large state and action space. 
Therefore, the computation is an underlying issue to be concerned. 
To handle it,   we establish two strategies to reduce the computation of   GB-L for HVAC control by studying the problem features. 

\emph{Strategy I: Concentrate on the high probability states.}  This strategy is promoted by the data analysis  (see Section IV-A). We observe though the outdoor temperature and humidity spread wide ranges  throughout  the day (i.e., temperature $[22, 34]$\si{\degreeCelsius} and relative humidity $[40, 100]$\si{\percent}), their variations over each specific time period are quite small.  For instance,  the outdoor temperature and humidity  mostly  lie  in the range of  $[26, 28]$\si{\degreeCelsius} and $[75, 95]$\si{\percent}) at  6:00 am. This suggests us to concentrate our computation on the states within those ranges while learning as the states outside are rare  and the actions have less impact on the overall performance.

\emph{Strategy II: Pick a good initial policy $\bm{\sigma}^0$}.   The operation of HVAC system should satisfy indoor comfort (avoid the occurrence of discomfort states). For HVAC control, we have some insight about the comfortable temperature and humidity ranges. 
For example, the temperature out of $[23, 28]^{\circ}$C or humidity out of  $[40, 70]\%$ will  cause discomfort (we can evaluate the PMV metrics).  Therefore, instead of a random pick, we can have a more reasonable initial policy $\bm{\sigma}^0$ by assigning the entries corresponding to the indoor temperature and humidity out of the ranges as \emph{zero}.  This is expected to  reduce the iterations. 

\vspace{-4mm}
\section{Applications}
In this section, we evaluate the performance of the proposed method for HVAC control via simulations. 
We organize this section into two parts.  \emph{First}, we  study the characteristics and establish the Markov chains of weather based on the meteorological data in Singapore.   \emph{Second},  We compare the method with the  existing MPC-based relaxation (MPC-R) method on the energy saving performance, thermal comfort, and on-line computing cost.

\vspace{-4mm}
\subsection{Data Analysis}

\begin{figure}[htbp]
	\setlength{\abovecaptionskip}{-2pt}
	\setlength{\belowcaptionskip}{2pt}
	\centering
	\subfigure[]{
		\begin{minipage}[t]{0.5\linewidth}
			\centering
			\includegraphics[width=1.6 in, height=1.2 in]{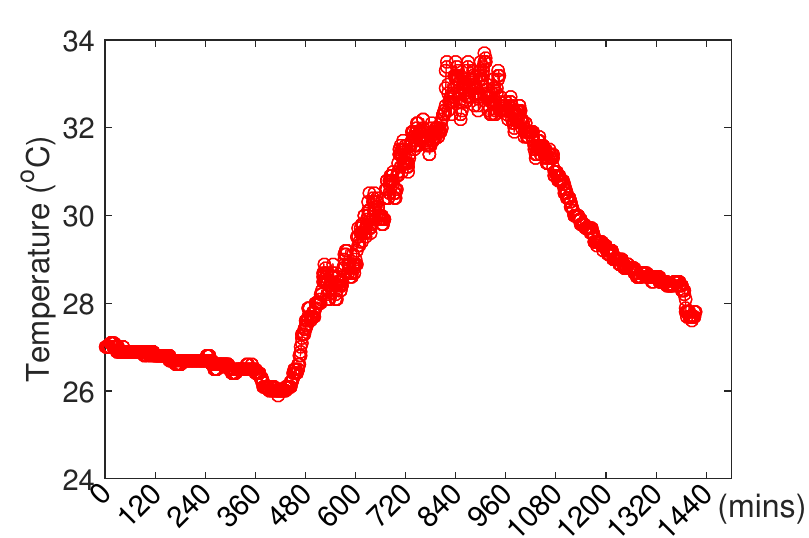}
		\end{minipage}%
	}%
	\subfigure[]{
		\begin{minipage}[t]{0.5\linewidth}
			\centering
			\includegraphics[width=1.6 in, height=1.2 in]{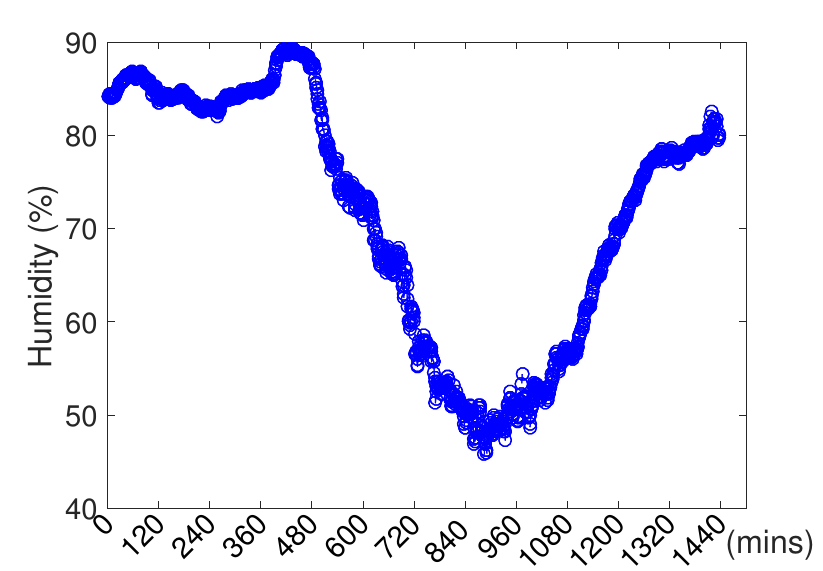}
		\end{minipage}%
	}%
	\centering
	\caption{ (a) Outdoor temperature for a typical day.  (b) Outdoor relative humidity for a typical  day.
	} \label{OutdoorWeather} 
\end{figure}

This section studies the characteristic of weather  based on  the meteorological data of Singapore (from 2019/09/01 to 2019/10/13,  43 days, minute resolution).
As an example,  we plot the temperature and humidity curve  for a typical day (2019/09/24)  in Fig. \ref{OutdoorWeather}. We can figure out some characteristics regarding the weather in tropical countries. For example, the temperature and (relative) humidity are   generally within the ranges of $[22, 34]$\si{\degreeCelsius} and  $[40, 100]$\si{\percent}.  Besides, the outdoor temperature is  usually  low in the early morning and gradually rise to the peak level at noon. After some point it begin to drop and reach the lowest level at late night. 
However, the  humidity shows the opposite patterns.

Based on the data, we establish two Markov chains to capture the dynamics of outdoor temperature and (relative) humidity.  It mainly includes two steps: \emph{i)} discretize the temperature and humidity data with a resolution of $1$\si{\degreeCelsius} and $5$ \si{\percent}, and \emph{ii)}  estimate the transition probabilities  by

\vspace{-3mm}
{\footnotesize
\begin{equation*}
\begin{split}
&\textrm{Temperature:}~\bm{P}^{\text T}_t[i, j] \approx \frac{{\sum_{\omega=1}^{ D} I (T^{\text{o}, \omega}_t=i, T^{\text{o}, \omega}_{t+1}=j)}}{\sum_{\omega=1}^{ D} I(T^{\text{o}, \omega}_t=i)}, \\
&\quad \quad \quad ~\forall i, j \in \{1, 2, \cdots, L_{\text T}\}, ~t\in\{1, \cdots, T-1\}.\\
&\textrm{Humidity:}~~\bm{P}^{\text H}_t [i, j]\approx \frac{{\sum_{\omega=1}^{ D} I(H^{\text{o}, \omega}_t=i, H^{\text{o}, \omega}_{t+1}=j)}}{\sum_{\omega=1}^{D} I(H^{\text{o},\omega}_t=i)}, \\
&\quad \quad \quad ~\forall i, j \in \{1, 2, \cdots, L_{\text H}\}, ~t\in\{ 1, \cdots, T-1\}.
\end{split}
\end{equation*}}
where $I(\cdot)$ is the indicator function. $\omega$ denotes the index of  the day,  and we have $D=43$.

\begin{figure}[htbp]
	\vspace{-5mm}
	\setlength{\abovecaptionskip}{-2pt}
	\setlength{\belowcaptionskip}{2pt}
	\centering
	\subfigure[]{
		\begin{minipage}[t]{0.5\linewidth}
			\centering
			\includegraphics[width=1.6 in, height=1.2 in]{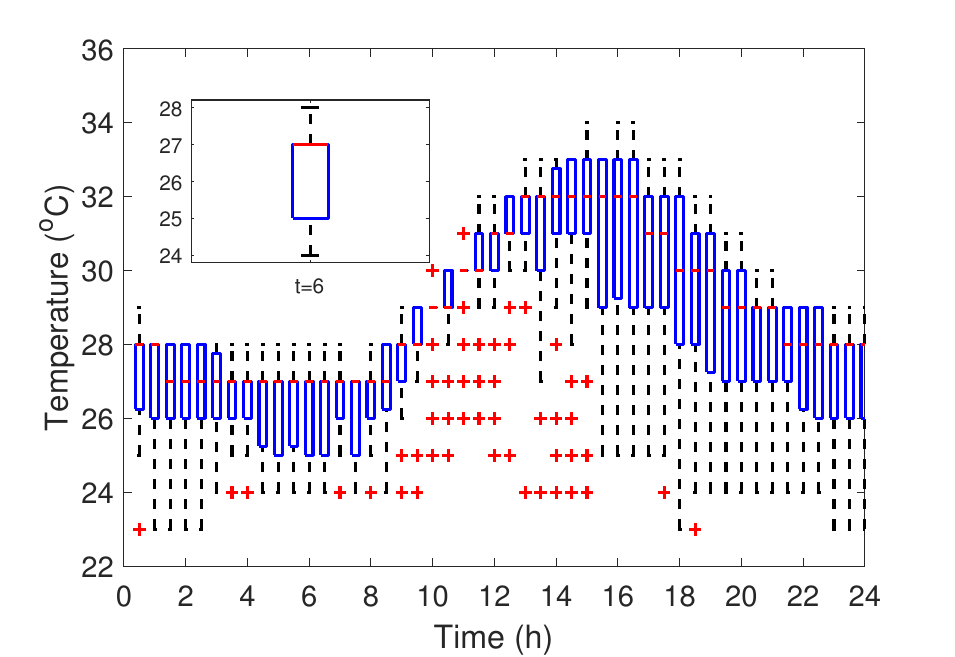}
		\end{minipage}%
	}%
	\subfigure[]{
		\begin{minipage}[t]{0.5\linewidth}
			\centering
			\includegraphics[width=1.6 in, height=1.2 in]{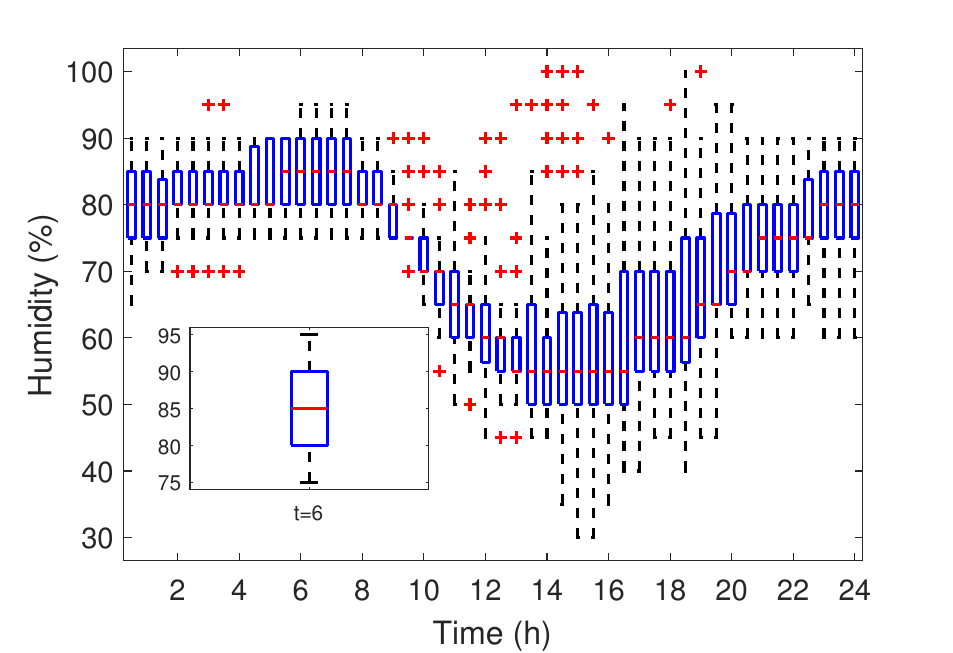}
		\end{minipage}%
	}%
	\centering
	\caption{ (a) The outdoor temperature distribution over the day.  (b) The outdoor relative humidity distribution over the day. } \label{WeatherDistribution} 
	\vspace{-6mm}
\end{figure}

Besides, we perform some analysis on the distribution of the outdoor temperature and  humidity during each hour period.
As shown in Fig. \ref{WeatherDistribution}, we observe though the outdoor temperature and humidity spread wide ranges  throughout  the day (i.e., temperature $[22, 34]$\si{\degreeCelsius} and relative humidity $[40, 100]$\si{\percent}), their variations over each specific time period are quite small.  For instance,  the outdoor temperature and humidity  mostly  lie  in the range of  $[26, 28]$\si{\degreeCelsius} and $[75, 95]$\si{\percent}) at  6:00 am.  This  indeed suggests the \emph{Strategy II} for computation reduction.

\subsection{Case Studies}
\emph{Simulation settings:} We consider an office of size $6$\si{\milli}$\times 5$\si{\milli}$\times 4$\si{\milli} and occupied by $5$ staffs. We equally divide the occupancy into $L_A= 6$ levels, corresponding to $0, 1, \cdots, 5$ occupants. The static inputs of the PMV model refer  to the ANSI/ASHRAE Standard \cite{ASHRAE55}. 
 The structural  parameters of the office  and  the heat/humidity generation  of the occupants are presented in TABLE \ref{RoomSettings} (referring to \cite{sun2013building}).    The nominal parameters of the HVAC system are covered in TABLE \ref{HVACParameters}. We use the time-of-use (TOU) price in Singapore to evaluate HVAC energy cost~ \cite{xu2017pmv}.
 \begin{table}[htbp]
 	\vspace{-5mm}
	\setlength{\abovecaptionskip}{-2pt}
	\setlength{\belowcaptionskip}{2pt}
	\scriptsize
	\centering
	\caption{Room \& Occupant Settings} \label{RoomSettings}
	\begin{tabular}{llll}
		\toprule[1.5pt]
		Param.    & Value \& Units &Param. & Value \& Units\\
		\hline
		$C_{\text p}$     &  $1012$\si{\joule \per {(\kilogram \cdot \kelvin)}}        &             $m^{\text a}$     & $144.6$\si{\kilogram}     \\
		$h_{\text{gs}}$  &  $2.5$\si{\watt \per{\milli^2}}                                   &   		          $m^{\text{wl}}$   & $7.2 \times 10^3$\si{\kilogram}\\
		$A_{\text{gs}}$  &  $10$\si{\milli^2}                                                      & 	            $m^{\text{wr}}$   & $8.64\times 10^3$\si{\kilogram}   \\
		$h_{\text w}$     & $0.8$\si{\watt \per{\milli^2}}                                     &	              $C_{\text w}$  & $1.05\times 10^3$\si{\joule \per {(\kilogram \cdot \kelvin)}} \\
		$a_{\text w}$     & $0.4$                                                                            &  	   $Q^{\text o}$     & $40$ \si{\joule \per \second}\\    
		$A_{\text{wl}}$  & $20$ \si{\milli^2}                                                      & 	       $H^{\text g}$    & $0.03$ \si{\gram\per \second}\\
		   	$A_{\text{wr}}$ & $24$\si{\milli^2}    \\
		\bottomrule[1.5pt]
	\end{tabular}
\end{table}

 \begin{table}[htbp]
 	\vspace{-5mm}
	\setlength{\abovecaptionskip}{-2pt}
	\setlength{\belowcaptionskip}{2pt}
	\scriptsize
	\centering
	\caption{HVAC \& PMV Parameters} \label{HVACParameters}
	\begin{tabular}{ll|ll}
		\toprule[1.5pt]
		\multicolumn{2}{c|}{HVAC} & \multicolumn{2}{c}{PMV} \\
		\hline
		Param.    & Val. \& Units &Param. & Val. \& Units\\
		\hline
		$F^{\text{fau},  \text{r}}$   & $0.1$ \si{\kelvin \watt}                      &  $v^{\text a}$       &   $0.2$\si{\milli \per {\second}}   \\
	    $F^{\text{fau},  \text{r}}$   & $0.1$ \si{\kelvin \watt}                      &   $M$                     &  $1.0$ met \\
		$G^{\text{fau}, \text{r}}$   & $0.01$ \si{\kilogram\per\second}   &  $W$                      &  $0$ \\
	    $G^{\text{fcu}, \text{r}}$   &  $0.05$ \si{\kilogram \per \second} &  $I_{\text{cl}}$    & $0.155$ clo\\
		$\eta$                                     &  $2.7$                                                      &   $P_{\text a}$     &  $1.01 \times 10^5$ \si{\pascal} \\
		\bottomrule[1.5pt]
	\end{tabular}
\vspace{-5mm}
\end{table}

Considering most existing on-line HVAC control  are deployed in MPC, we compare the GB-L with an MPC method. Particularly, to identify the performance bounds, we assume perfect information (i.e., weather and occupancy) for the MPC in  our simulations, which is usually not available in realization but  to create  a benchmark here.  
As for the MPC method  with a receding horizon $H$,  we have  the formulation:

\vspace{-3mm}
\small
\begin{eqnarray} \label{MPC method}
\begin{split}
&\quad \quad \min_{\substack{G^{\textrm{fau}}_t,G^{\textrm{fcu}}_t, \\
		T^{\textrm{fau}}_t,  T^{\textrm{fcu}}_t}} J(t_k)=\!\!\!\sum_{\mathclap{t=t_k}}^{t_k+H-1}\!\!\! c_t \Big\{  \eta (C^{\textrm{fcu}}_t \!+\!C^{\textrm{fau}}_t)  +\!F^{\textrm{fcu}}_t\!+F^{\textrm{fau}}_t \Big\}\Delta_t \\
&s.t.~~ \textrm{{System dynamics}} : (\ref{indoor temperature})-(\ref{humidity}), ~ \textrm{{Operation limits:} (\ref{air flow rate constraints}}),\\
&\quad~~~\textrm{{Thermal comfort:~}} (\ref{thermal comfort}), ~\forall t\in \{ t_k, t_k+1, \cdots, t_k\!+\!H\!-\!1\}.\\
\end{split}
\end{eqnarray}

\normalsize
As discussed,  solving problem \eqref{MPC method} exactly is a nontrivial task due to the non-linear  and non-analytical constraints imposed by the system dynamics and the PMV model. 
To build the comparisons,  we  settle for  an existing  relaxed solution method (referred to MPC-R) \cite{xu2015supply, xu2017pmv}  as it is  the scarcely available and capable one for problem \eqref{MPC method} to our best knowledge. The main point of MPC-R is to convert the non-linear constraints into a sequence of combinatorial linear constraints through piece-wise linearization and approximation o as to be  tackled by some commercial solvers, like CPLEX.  

We compare the two methods under three different settings regarding the state and control discretization. Particularly, the  state discretization  is only  required by GB-L not the MPC-R. 

\textbf{S-1}: ~The discretization pace of  temperature and  relative humidity are $2$\si{\degreeCelsius} and $10$\si{\percent}, respectively. The set-point temperature levels  for  FAU and FCU are  $\{12, 14, 16\}$\si{\degreeCelsius}, and their supply air flow rates are equally divided into $3$ levels.

\textbf{S-2}:~ The discretization of temperature  and relative humidity are  as \textbf{S-1}. The set-point temperature of FAU and FCU are fixed as  $15$\si{\degreeCelsius}, and their supply air flow rates are equally divided into $5$ levels.

\textbf{S-3}:~The discretization pace of  temperature  and  relative humidity are  $1$\si{\degreeCelsius} and  $5$\si{\percent}. The settings for  FAU and FCU  are  as \textbf{S-2}.

\begin{figure}
	\setlength{\abovecaptionskip}{-2pt}
	\setlength{\belowcaptionskip}{2pt}
	\centering
	\includegraphics[width=3.0 in, height = 1.2 in,  keepaspectratio ]{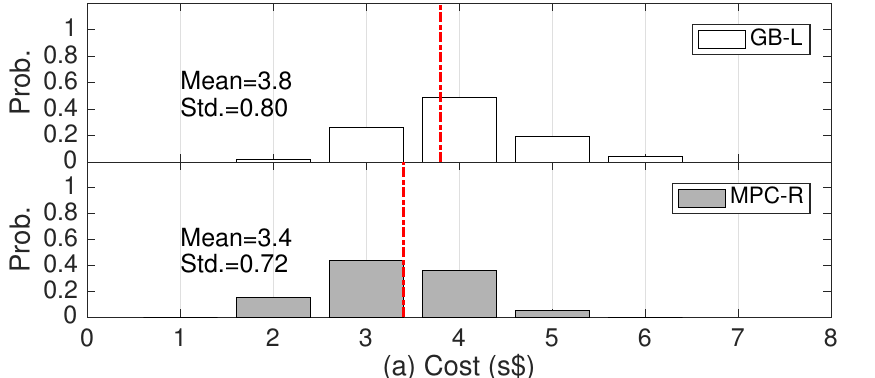}
	\includegraphics[width=3.0 in, height = 1.2 in,  keepaspectratio]{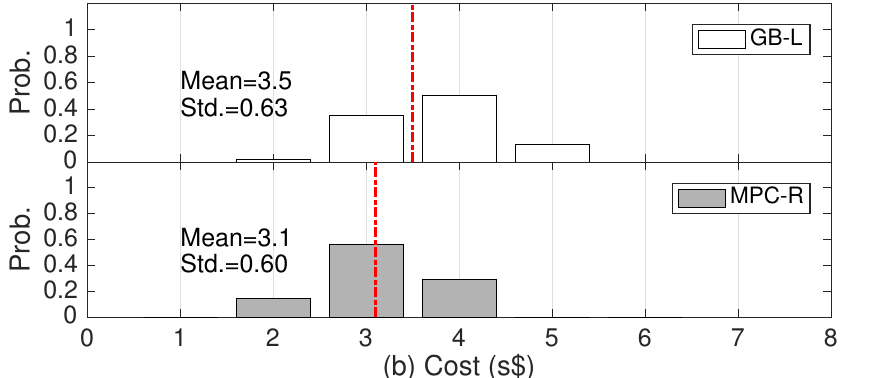}\\
	\includegraphics[width = 3.0 in,height = 1.2 in,  keepaspectratio]{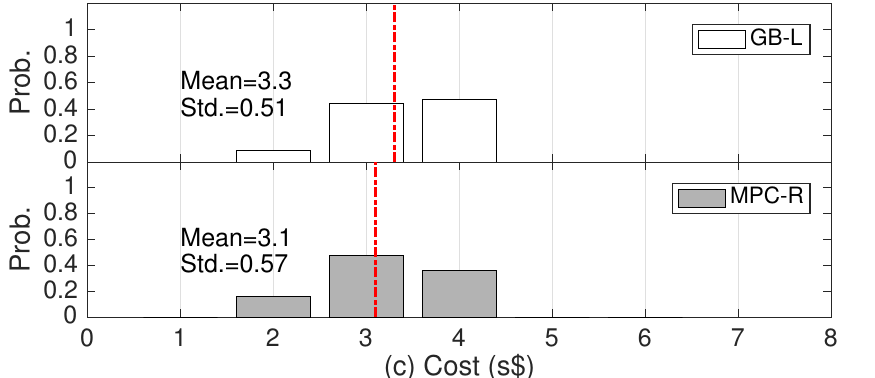}\\
	\caption{The histograms of HVAC costs: (a) \textbf{S-1}. (b) \textbf{S-2}.  (c) \textbf{S-3}.}\label{HVAC_Cost}
	\vspace{-4mm}
\end{figure}

For the GB-L, we use $1000$ (\textbf{S-1}), $2000$ (\textbf{S-2}) and $5000$ (\textbf{S-3}) sample paths to estimate the performance gradients while learning the policy. 
As there exist uncertainties, we compare the performance (i.e., energy cost, thermal comfort and on-line computation cost) of  two methods under $100$ randomly generated scenarios. First of all, we study the distributions of  induced energy cost by the two methods as shown in  Fig. \ref{HVAC_Cost}.  
We see some  performance discount  of  GB-L over MPC-R:  $11.7$\si{\percent} (\textbf{S-1}), $12.9$\si{\percent} (\textbf{S-2}) and $6.5$\si{\percent} (\textbf{S-3}), respectively.   This is reasonable as we have assumed  accurate information for the MPC-R to provide the  near-optimal bounds. 
Besides,  the variation of performance of the GB-L with the different settings can be attributed to: \emph{i)} the state discretization pace,   and \emph{ii)} the estimation accuracy of performance gradients affected by the number of sample paths used. 
Reasonably, we can expect to shrink the performance gap further with a finer-grained state discretization and increased sample paths  if more computation is acceptable. This is inferred from  the comparison of   \textbf{S-2} and \textbf{S-3}.



Further, we study the on-line computation cost for the two methods. Particularly, GB-L learns the optimal control policy off-line  and stores it for on-line execution. Therefore, the main on-line computation lies in identify the control from the state-action mapping table. Whereas the implementation of MPC-R requires to solve problem \eqref{MPC method} based on the updated information at each stage. For the different settings,  the mean and standard deviation (std.) of on-line computation  \footnote{ Simulation platform: Matlab R2016a, Window i7-5500 CPU@2.40GHZ. } throughout  the optimization horizon  are contrasted in TABLE \ref{on-line-computation}.  We see the GB-L can respond instantly ( less than 1 sec.),   whereas the MPC-R  takes 4-5  minutes.   This demonstrates GB-L can enable efficient on-line implementation.
\begin{table}[h] \label{on-line-computation}
	\vspace{-5mm}
	\setlength\tabcolsep{3pt}
	\centering
	\caption{On-line computing time of GB-L and MPC-R.}
	\label{tab:SimulationTime}
	\begin{tabular}{ccccc}
		\toprule 
		\multirow{2}{*}{\#Settings~~}&  \multicolumn{2}{c}{\textbf{GB-L}} &   \multicolumn{2}{c}{\textbf{MPC-R} }  \\
		& Mean (sec.)~  & Std (sec.) ~~~            &~~~ Mean (sec.)~  & Std(sec.)          \\
		\hline 
		\textbf{S-1}   &   0.48          & 0.09          &  259.83    &  40.82    \\
		\textbf{S-2}   &   0.58          & 0.13          &  247.41    &  41.91  \\
		\textbf{S-3}   &     0.43        &  0.06        &  335.39    &   43.14\\
		\bottomrule 
	\end{tabular}
\end{table} 

\begin{figure}[h]
	\setlength{\abovecaptionskip}{-2pt}
	\setlength{\belowcaptionskip}{2pt}
	\centering
	\includegraphics[width=3.0 in, height=1.7 in,  keepaspectratio ]{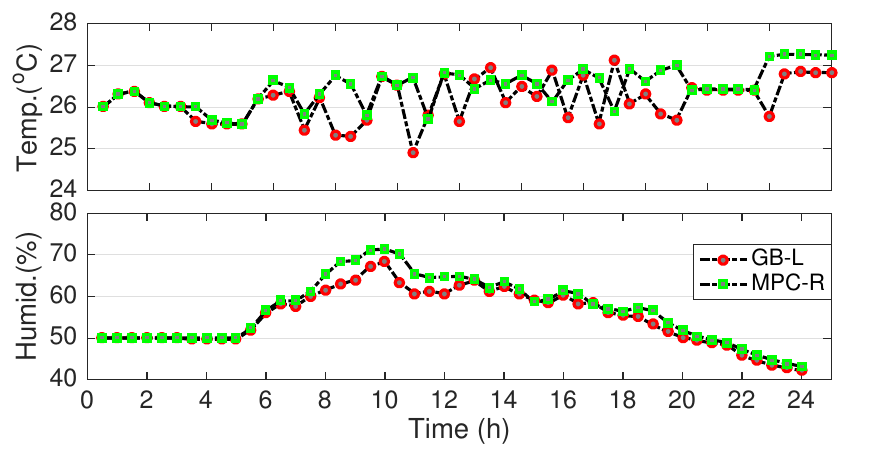}\\
	\caption{The indoor temperature (Temp.) and relative humidity (Humid.) for a specific scenario with the GB-L and MPC-R.}\label{TemperatureHumidity}
\end{figure}

\begin{figure}[h]
	\setlength{\abovecaptionskip}{-2pt}
	\setlength{\belowcaptionskip}{2pt}
	\centering
	\includegraphics[width=3.0 in, height=1.6 in,  keepaspectratio]{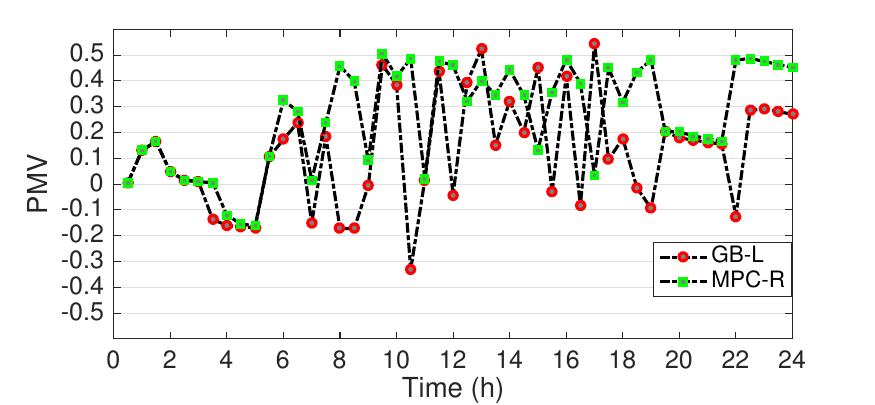}\\
	\caption{The PMV curves for a scenario with the GB-L and MPC-R method.}\label{PMV curves}
\end{figure}

\begin{figure}[h]
	\vspace{-3mm}
	\setlength{\abovecaptionskip}{-2pt}
	\setlength{\belowcaptionskip}{2pt}
	\centering
	\includegraphics[width=3.2 in, height=1.2 in,  keepaspectratio]{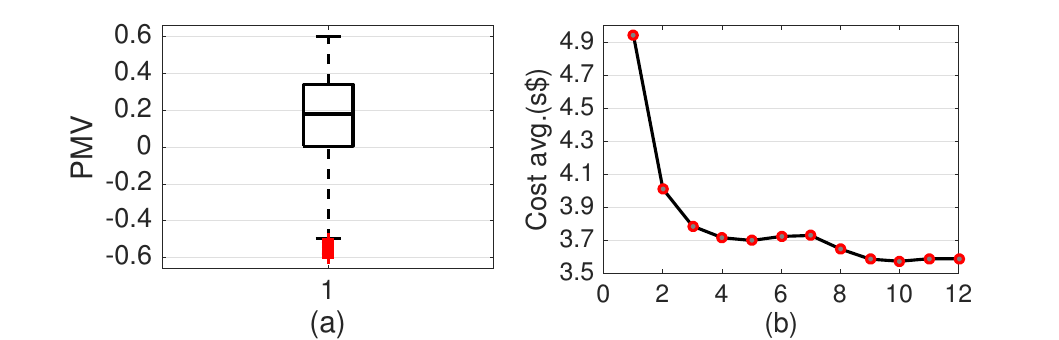}\\
	\caption{(a) The distribution of PMV value with the GB-L (\textbf{S-2}). (b) The convergence rate of GB-L (\textbf{S-2}).}\label{PMV distribution}
\end{figure}

Subsequently, we evaluate the thermal comfort under the two methods  by inspecting
the indoor temperature, humidity and the PMV value for a randomly picked scenario with \textbf{S-2}.
As show  in Fig. \ref{TemperatureHumidity}, we observe both the indoor temperature and  relative humidity are maintained in the typical comfortable range $[24, 27]$\si{\degreeCelsius} and $[40, 70]$\si{\percent} with the two methods.  More notably, 
as  shown in Fig. \ref{PMV curves}, the PMV curve lies in the specified  range of $[-0.5, 0.5]$.   
However, to be noted is that we may observe  comfort violations with other realizations due the uncertainties. To evaluate the  thermal comfort with GB-L under uncertainties,  
we study the distribution of PMV value under the $100$ realizations. 
As indicated in Fig. \ref{PMV distribution} (a),  the PMV value is almost maintained within the range of  $[-0.5, 0.6]$. Particularly,  It's reasonable to see a minor upper violation (i.e., 0.6) as  the energy  cost saving target is set. Further, we find the exact thermal comfort i.e., PMV $[-0.5, 0.5]$ is achieved with a probability of $93\%$. This demonstrates the desirable performance of the GB-L for providing thermal comfort under uncertainties. 

\begin{figure}[h]
	\setlength{\abovecaptionskip}{-2pt}
	\setlength{\belowcaptionskip}{2pt}
	\centering
	\includegraphics[height = 1.2 in,width=2.8 in,  keepaspectratio]{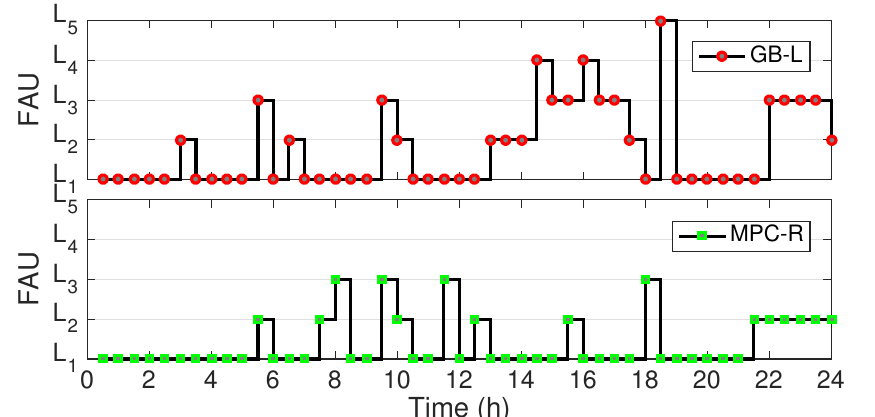}\\
	\includegraphics[height = 1.2 in, width=2.8 in,  keepaspectratio]{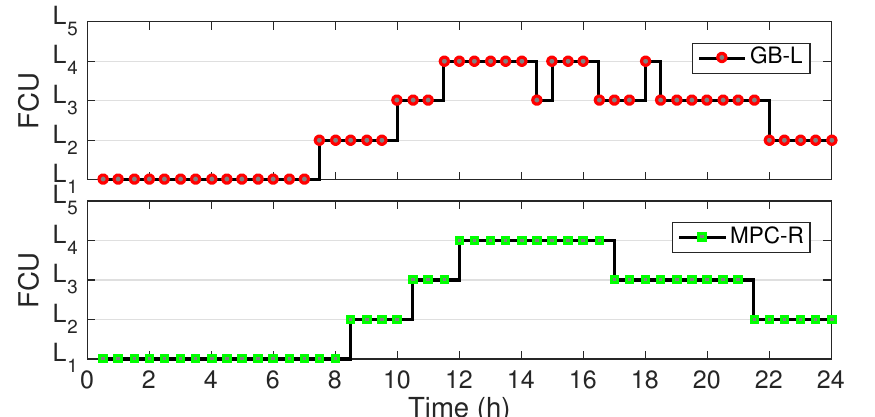}
	\caption{(a) The control of FAU and FCU  for a specific scenario with the GB-L and MPC-R method ($L_1$-$L_5$ denotes the air flow rate levels).}\label{FAU_FCUControl}
\end{figure} 

\begin{figure}
	\setlength{\abovecaptionskip}{-2pt}
	\setlength{\belowcaptionskip}{2pt}
	\centering
	\includegraphics[clip, trim=2cm 8.7cm 2cm 8.2cm, width=0.40\textwidth]{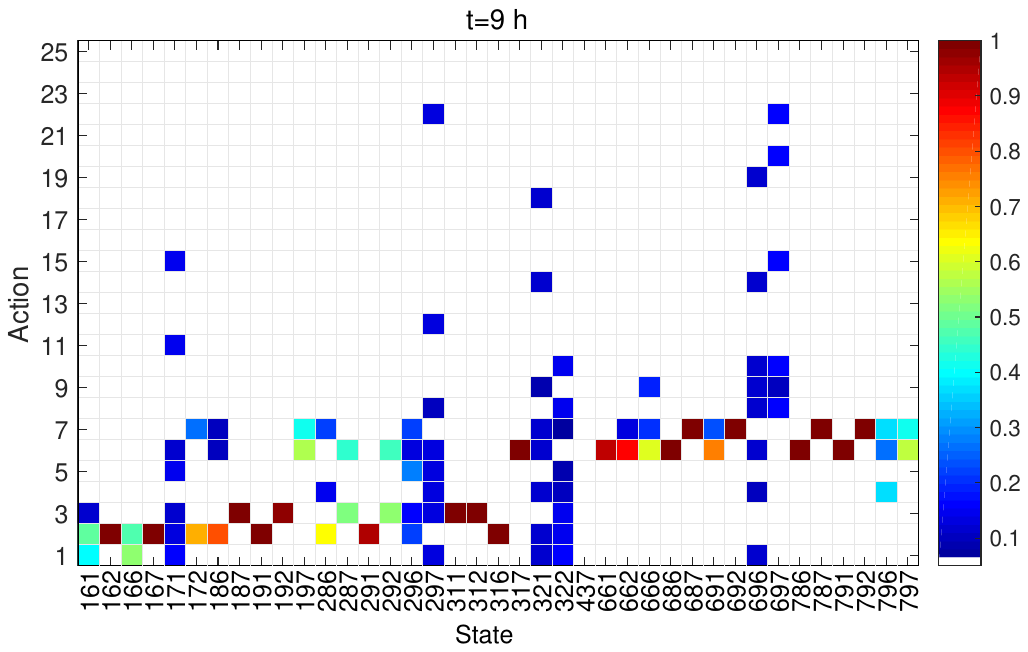} \\
	\includegraphics[clip, trim=2cm 8.7cm 2cm 8.2cm, width=0.40\textwidth]{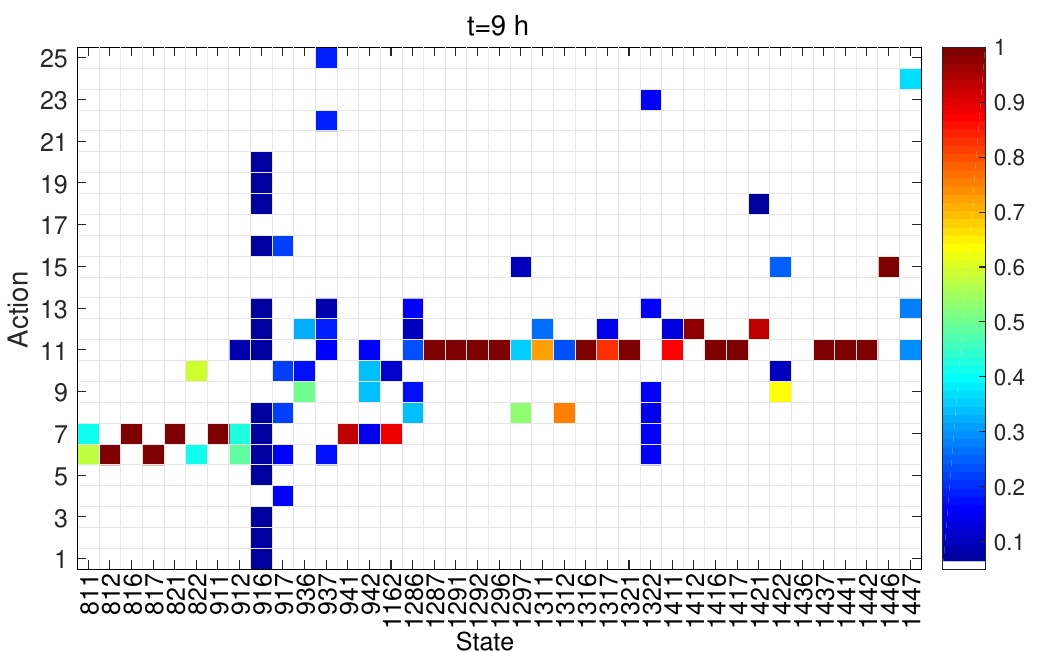} \\
	\caption{The stochastic policy obtained by GB-L for 9:00 a.m. (\textbf{S-2}).}\label{Randompolicy1}
\end{figure}

Use this specific scenario as an example, we also study the control inputs for the FAU and FCU.
As shown in \ref{FAU_FCUControl},  we see quite similar operation patterns of FAU/FCU  with  the two methods. 
This implicitly illustrates the energy saving  performance of GB-L that we observed in Fig. \ref{HVAC_Cost}.
Besides, we can observe some other interesting phenomenon that  the FAU are mostly operated in quite lower level versus the FCU.
This  is  rational  as the FAU handles outdoor fresh air with a general higher temperature than that of the recirculated air managed by the FCU. To fulfill the energy saving objective, the HVAC system tends to activate the FCU instead of FAU in priority to reduce cooling load. 
Besides,  we observe  the operation patterns, especially the FCU,   correspond well to the typical weather patterns (i.e., lower outdoor temperature in the early morning and late night,  and higher temperature in the noon)  and the occupancy patterns (i.e., high occupancy during the working hours and low occupancy during non-working hours).  The demonstrates the capability of GB-L responding to the uncertain  thermal disturbance caused  by the weather and occupancy.

Though the GB-L is  implemented off-line,  the convergence speed is still an important and concerned issue.  
Therefore, we inspect the convergence rate of the GB-L with \textbf{S-2}.  We inspect the average energy cost of $100$ realizations while executing  \textbf{Algorithm} \ref{GB-L}. 
Take the average energy cost as  an indicator, we obtain the convergence rate in Fig. \ref{PMV distribution} (b).  
For this case,  it takes about $10$ iterations to approach the ``optima".

\begin{figure}
	\centering
	\includegraphics[width=3.0 in]{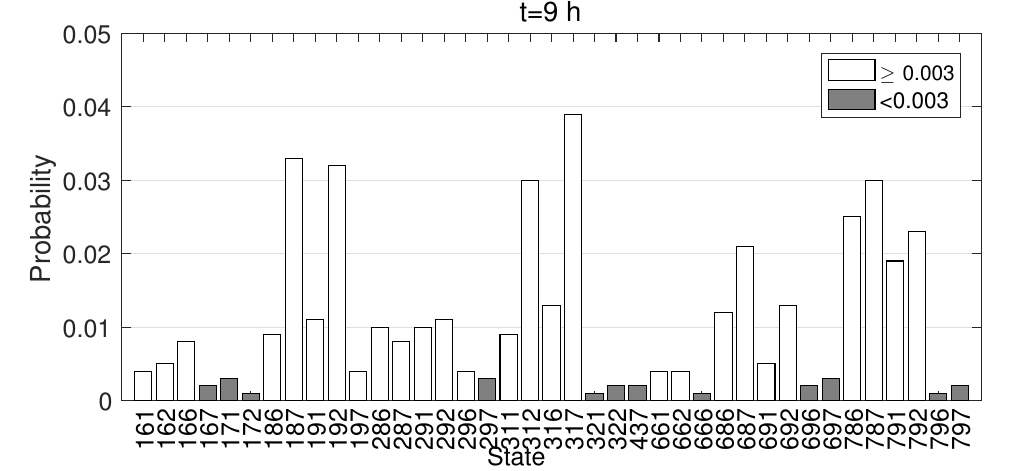}\\
		\includegraphics[width=3.0 in]{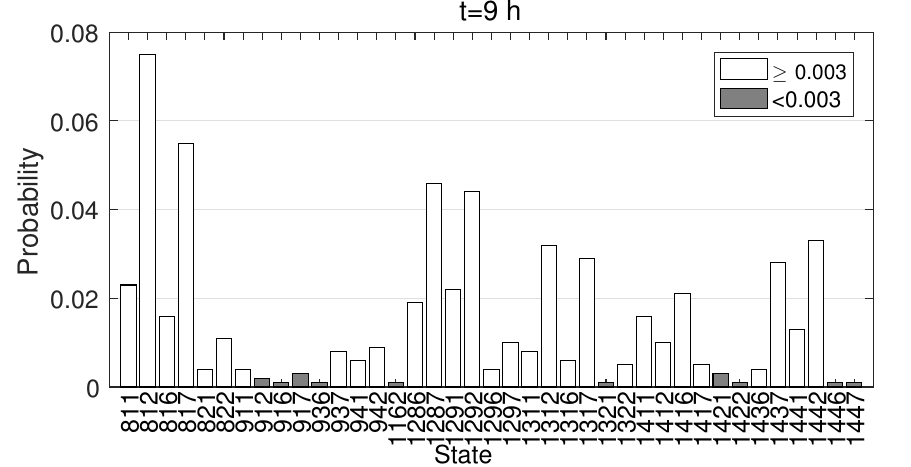}\\
	\caption{The state distribution at 9:00 a.m. \textbf{S-2}.}\label{StateDistribution9am1}
	\vspace{-4mm}
\end{figure}

Last, we visualize the obtained  stochastic policy (a mapping from the state space to the action space) at 9:00 a.m. in Fig. \ref{Randompolicy1} (only for the visited states in the sample paths) as an example.  From the figure, we can have some insights regarding the
characteristics of GB-L for HVAC control.  We note that for many states, the probability is concentrated within a  small groups of actions.  However, for some states like  $\{171, 296, 297, 321, 322, 696, 697, \cdots \}$, the probability distributions   scatter more diversely on  the action space. The latter is attributed to the low occurrence of  those states in the sample paths (see \ref{StateDistribution9am1}), resulting in fewer updates.  Conversely, this implies those states occur with  lower probability   and deserve less computation.

\section{Conclusion}
This paper studied the energy-efficient control of HVAC systems  subject to uncertain thermal disturbances caused by the weather and the occupants. To enhance thermal comfort, we incorporate the elaborate predictive mean vote (PMV) thermal comfort model in the optimization. This problem suffers computational challenges from the the non-linear and non-analytical constraints imposed by the thermal dynamics and  PMV model. 
To handle it, we formulated the problem as an Markov decision process (MDP) and proposed a gradient-based learning (GB-L) method for learning the optimal stochastic control policy off-line and stored it for efficient on-line implementation. We prove  the method’s converge to the optimal policies theoretically. In parallel, we demonstrated the method’s performance (i.e., energy cost, thermal comfort and on-line computing) for HVAC control via simulations. The comparisons with the existing MPC-based relaxation (MPC-R) method which was assumed with accurate future information and expected to provide the near-optimal bounds, 
showed that though there exists some performance loss in energy cost savings, the proposed method can enable efficient on-line implementation and provide high probability of thermal comfort under uncertainties. 

This paper has used singe office as an example to establish the GB-L method. An interesting future direction is its extension to multi-zone commercial buildings. However, that's not straightforward due to the concatenated state and action space to handle.  From our  insight, one possible solution is to adopt  the \emph{``one-agent-at-a-time"} idea ( i.e, sequentially learning the control policies for each individual zones) proposed in \cite{bertsekas2019multiagent} to address  the computational issue.

\vspace{-3mm}
\appendices
\section{Proof of \textbf{Theorem} \ref{theorem1}}
\vspace{-2mm}
\begin{proof}
According to (\ref{performance difference}),  we have the following performance difference equation over two successive iterations:
\small
\begin{equation}\label{(31)}
\begin{split}
&J(\bm{\sigma}^{k+1}; S_0)-J(\bm{\sigma}^{k}; S_0)\\
&\!=\!\sum_{t=0}^{T-1} \pi^{\bm{\bm{\sigma}^{k+1}}}_t \Big[ (\bm{r}^{\bm{\sigma}^{k+1}}_t\!\!-\!\!\bm{r}^{\bm{\sigma}^k}_t)\!+\!(\bm{P}^{\bm{\sigma}^{k+1}}_t\!\!-\!\!\bm{P}^{\bm{\sigma}^k}_t)\bm{V}^{\bm{\sigma}^k}_{t+1}\Big]\\
&=\sum_{t=0}^{T-1} \sum_{\bm{s}_t\in \mathcal{S}_t }\pi^{\bm{\bm{\sigma}^{k+1}}}_t(\bm{s}_t) \Big[ (\bm{r}^{\bm{\sigma}^{k+1}}_t(\bm{s}_t)-\bm{r}^{\bm{\sigma}^k}_t(\bm{s}_t))\\
&\quad \quad +\sum_{\mathclap{\bm{s}_{t+1}\in\mathcal{S}_{t+1}}}(\bm{P}^{\bm{\sigma}^{k+1}}_t(\bm{s}_{t+1}|\bm{s}_t)\!-\!\bm{P}^{\bm{\sigma}^k}_t(\bm{s}_{t+1}|\bm{s}_t))V^{\bm{\sigma}^k}_{t+1}(\bm{s}_{t+1})\Big]\\
\end{split}
\end{equation}
\normalsize
For brevity, we define an operator as $\Sigma^k(\bm{s}_t)=\sum_{\bm{a}_t=1}^{|\mathcal{A}_t|} \sigma^k_t(\bm{s}_t, \bm{a}_t)$ and we have
\begin{equation}\label{(32)}
\begin{split}
&r_t^{\bm{\sigma}^{k+1}}(\bm{s}_t)-r_t^{\bm{\sigma}^{k}}(\bm{s}_t)\!\\
&\quad =\!\!\sum_{\bm{b}_t=1}^{|\mathcal{A}_t|} \big[p_t^{\bm{\sigma}^{k+1}}(\bm{b}_t|\bm{s}_t)-p_t^{\bm{\sigma}^k}(\bm{b}_t|\bm{s}_t)\big] r_t(\bm{s}_t, \bm{b}_t)\\
&\quad =\sum_{\bm{b}_t=1}^{|\mathcal{A}_t|} \Big[\frac{\sigma_t^{k+1}(\bm{s}_t, \bm{b}_t)}{\Sigma^{k+1}(\bm{s}_t)}-\frac{\sigma_t^k(\bm{s}_t, \bm{b}_t)}{\Sigma^k(\bm{s}_t)}\Big] r_t(\bm{s}_t, \bm{b}_t)\\
\end{split}
\end{equation}
\begin{equation}\label{(33)}
\begin{split}
&\bm{P}^{\bm{\sigma}^{k+1}}_t(\bm{s}_{t+1}|\bm{s}_t)-\bm{P}^{\bm{\sigma}^k}_t(\bm{s}_{t+1}|\bm{s}_t)\\
&\quad =\sum_{\bm{b}_t=1}^{|\mathcal{A}_t|} p_t(\bm{s}_{t+1}|\bm{s}_t, \bm{b}_t) (p_t^{\bm{\sigma}^{k+1}}(\bm{b}_t|\bm{s}_t)-p_t^{\bm{\sigma}^k}(\bm{b}_t|\bm{s}_t))\\
&\quad =\sum_{\bm{b}_t=1}^{|\mathcal{A}_t|} p_t(\bm{s}_{t+1}|\bm{s}_t, \bm{b}_t)\Big[\frac{\sigma_t^{k+1}(\bm{s}_t, \bm{b}_t)}{\Sigma^{k+1}(\bm{s}_t)}-\frac{\sigma_t^k(\bm{s}_t, \bm{b}_t)}{\Sigma^k(\bm{s}_t)}\Big]
\end{split}
\end{equation}

By substituting (\ref{(32)}) and (\ref{(33)})  into (\ref{(31)}), we have
\small
\begin{equation} \label{(34)}
\begin{split}
J&(\bm{\sigma}^{k+1}; S_0)-J(\bm{\sigma}^{k}; S_0)\\
&=\!\sum_{t=0}^{T-1} \sum_{\bm{s}_t\in \mathcal{S}_t } \Big[ \sum_{\bm{b}_t=1}^{|\mathcal{A}_t|} [\frac{\sigma_t^{k+1}(\bm{s}_t, \bm{b}_t)}{\Sigma^{k+1}(\bm{s}_t)}\\
&\quad \quad \quad \quad \quad-\!\frac{\sigma_t^k(\bm{s}_t, \bm{b}_t)}{\Sigma^k(\bm{s}_t)}] \big(r_t(\bm{s}_t, \bm{b}_t)+ V^{\bm{\sigma}^k}_{t}(\bm{s}_t, \bm{b}_t)\big)\Big]\\
&=\!\sum_{t=0}^{T-1} \sum_{ \bm{s}_t\in \mathcal{S}_t } \sum_{ \bm{b}_t=1}^{|\mathcal{A}_t|} [\frac{\sigma_t^{k+1}(\bm{s}_t, \bm{b}_t)}{\Sigma^{k+1}(\bm{s}_t)}\!-\!\frac{\sigma_t^k(\bm{s}_t, \bm{b}_t)}{\Sigma^k(\bm{s}_t)}]A^{\bm{\sigma}^k}(\bm{s}_t, \bm{b}_t) 
\end{split}
\end{equation}
\normalsize
where we have $A_t^{\bm{\sigma}^k}(\bm{s}_t, \bm{b}_t)=r_t(\bm{s}_t, \bm{b}_t)+V_t^{\bm{\sigma}^k}(\bm{s}_t, \bm{b}_t)$.

As indicated in  (\ref{policy iteration}), we have
\begin{displaymath}
\begin{split}
\sigma^{k+1}(\bm{s}_t, \bm{a}_t)=\sigma_t^k(\bm{s}_t, \bm{a}_t)&-\gamma^k(\bm{s}_t, \bm{a}_t) \Delta \sigma_t^k(\bm{s}_t, \bm{a}_t), \\
       &~\forall \bm{a}_t \in \mathcal{A}_t.\\
\end{split}
\end{displaymath}
where we have 
\begin{equation*} \label{(36)}
\begin{split}
&\Delta \sigma_t^k( \bm{s}_t, \bm{a}_t)=\frac{\partial J(\bm{\sigma}^k; \bm{S}_0)}{\partial {\sigma}_t^k(\bm{s}_t, \bm{a}_t)}\\
&= \pi^{\bm{\sigma}^k}_t (\bm{s}_t) \Big[\sum_{\bm{b}_t=1}^{|\mathcal{A}_t|} \frac{\partial p^{\bm{\sigma}^k}(\bm{b}_t|\bm{s}_t)}{\partial \sigma_t^k(\bm{s}_t, \bm{a}_t)}(r_t(\bm{s}_t, \bm{b}_t)+V^{\bm{\sigma}^k}_{t}(\bm{s}_t, \bm{b}_t))\Big]\\
&=\pi^{\bm{\sigma}^k}_t (\bm{s}_t) \Big[ \sum_{\bm{b}_t=1, \bm{b}_t\neq \bm{a}_t} ^{|\mathcal{A}_t|}  \frac{-\sigma_t^k(\bm{s}_t, \bm{b}_t)}{[\Sigma^k(\bm{s}_t)]^2} A^{\bm{\sigma}^k}_t(\bm{s}_t, \bm{b}_t) \\
\end{split}
\end{equation*}
\begin{equation*} 
\begin{split}
&\quad \quad \quad \quad \!+\!\frac{\Sigma^k(\bm{s}_t)-\sigma_t^k(\bm{s}_t, \bm{a}_t)}{[\Sigma^k(\bm{s}_t)]^2} A_t^{\bm{\sigma}^k}(\bm{s}_t, \bm{a}_t)\Big]\\
&=\frac{\pi^{\bm{\sigma}^k}_t (\bm{s}_t) }{[\Sigma^k(\bm{s}_t)]^2}\Big[\Sigma^k (\bm{s}_t) A_t^{\bm{\sigma}^k}(\bm{s}_t, \bm{a}_t)- \sum_{\bm{b}_t=1}^{|\mathcal{A}_t|} \sigma^k(\bm{s}_t, \bm{b}_t) A_t^{\bm{\sigma}^k}(\bm{s}_t, \bm{b}_t)\Big]
\end{split}
\end{equation*}

Besides, regarding \eqref{(34)},  we have
\begin{equation} \label{(37)}
\begin{split}
&\frac{\sigma_t^{k+1}(\bm{s}_t, \bm{a}_t)}{\Sigma^{k+1}(\bm{s}_t)}\!-\!\frac{\sigma_t^k(\bm{s}_t, \bm{a}_t)}{\Sigma^k(\bm{s}_t )}\\
& = \frac{\Delta \sigma_t^k(\bm{s}_t, \bm{a}_t) \cdot \Sigma^k(\bm{s}_t)+  \Delta \Sigma^k (\bm{s}_t)\cdot \sigma_t^k(\bm{s}_t, \bm{a}_t)}{\Sigma^k(\bm{s}_t)\cdot \Sigma^{k+1}(\bm{s}_t)}\\
\end{split}
\end{equation}
where $\Delta \Sigma^k (\bm{s}_t)=\Sigma^{k+1} (\bm{s}_t)-\Sigma^k (\bm{s}_t)$.

From \eqref{(37)}, we observe that if the stepsizes are selected as $\gamma_t^k(\bm{s}_t, \bm{b}_t)=\frac{\sigma_t^k(\bm{s}_t, \bm{b}_t)}{\Sigma^k(\bm{s}_t)}$ ($\forall \bm{b}_t \in \{1, 2, $ $\cdots,  |\mathcal{A}_t|\}$), we  have 
\small
\begin{equation} \label{(38)}
\begin{split}
&\Delta \Sigma^k (\bm{s}_t)=\sum_{\bm{b}_t=1}^{|\mathcal{A}_t|} \bm{\gamma}_t^k(\bm{s}_t, \bm{b}_t) \Delta \sigma_t^k(\bm{s}_t, \bm{b}_t)\\
&=\frac{\pi^{\sigma^k}_t (\bm{s}_t) }{[\Sigma^k(\bm{s}_t)]^2} \Big[\sum_{\mathclap{\bm{b}_t=1}}^{|\mathcal{A}_t|} \gamma_t^k(\bm{s}_t, \bm{b}_t)\Sigma^k(\bm{s}_t)  A_t^{\bm{\sigma}^k}(\bm{s}_t, \bm{b}_t)\\
&\quad \!-\!\sum_{\bm{b}_t=1}^{|\mathcal{A}_t|} \gamma_t^k(\bm{s}_t, \bm{b}_t) \sum_{\bm{b}_t=1}^{|\mathcal{A}_t|} \sigma_t^k(\bm{s}_t, \bm{b}_t)A_t^{\bm{\sigma}^k}(\bm{s}_t, \bm{b}_t)\Big]\\
&=0
\end{split}
\end{equation}
\normalsize

The equation (\ref{(38)}) implies  $\Sigma^k(\bm{s}_t)=\Sigma^{k+1}(\bm{s}_t)$ ($\forall  \bm{s}_t\in \mathcal{S}_t $) with the selected stepsize $\gamma_t^k(\bm{s}_t, \bm{b}_t)=\frac{\sigma_t^k(\bm{s}_t, \bm{b}_t)}{\Sigma^k(\bm{s}_t)}$ ($\forall \bm{s}_t \in \mathcal{S}_t , \bm{b}_t \in  \mathcal{A}_t$).

By substituting (\ref{(38)}) into (\ref{(37)}), we have 
\begin{eqnarray} \label{(39)}
\begin{split}
&\frac{\sigma_t^{k+1}( \bm{s}_t, \bm{a}_t)}{\Sigma^{k+1}(\bm{s}_t)}-\frac{\sigma_t^k(\bm{s}_t, \bm{a}_t)}{\Sigma^k(\bm{s}_t)}= \frac{-\gamma_t^k (\bm{s}_t, \bm{a}_t) \Delta \sigma_t^k(\bm{s}_t, \bm{a}_t) }{\Sigma^{k}(\bm{s}_t)}\\
&= \frac{-\pi^{\bm{\sigma}^k}_t (\bm{s}_t) }{[\Sigma^k(\bm{s}_t)]^3}\Big[\Sigma^k(\bm{s}_t)\gamma_t^k (\bm{s}_t, \bm{a}_t)A_t^{\bm{\sigma}^k}(\bm{s}_t, \bm{a}_t)\\
&\quad \quad \quad  \!-\!\gamma_t^k(\bm{s}_t, \bm{a}_t) \sum_{\bm{b}_t=1}^{|\mathcal{A}_t|} \sigma^k(\bm{s}_t,  \bm{b}_t)A_t^{\bm{\sigma}^k}(\bm{s}_t, \bm{b}_t)\Big]
\end{split}
\end{eqnarray}

Then by substituting (\ref{(39)}) into (\ref{(34)}) we have
\small
\begin{eqnarray}
\begin{split}
&J(\bm{\sigma}^{k+1}; S_0)-J(\bm{\sigma}^{k}; S_0)\\
&\!=\!\sum_{t=0}^{T-1}\sum_{\bm{s}_t\in \mathcal{S}_t }\sum_{\bm{a}_t=1}^{|\mathcal{A}_t|} \Big\{\frac{-\pi^{\bm{\sigma}^k}_t (\bm{s}_t) }{[\Sigma^k(\bm{s}_t)]^3}\Big[\Sigma^k(\bm{s}_t)\gamma_t^k (\bm{s}_t, \bm{a}_t)A_t^{\bm{\sigma}^k}(\bm{s}_t, \bm{a}_t) \\
&\quad \quad \!-\!\gamma_t^k(\bm{s}_t, \bm{a}_t) \sum_{\bm{a}_t=1}^{|\mathcal{A}_t|} \sigma_t^k( \bm{s}_t, \bm{a}_t)A_t^{\bm{\sigma}^k}(\bm{s}_t, \bm{a}_t)\Big] A_t^{\bm{\sigma}^k}( \bm{s}_t, \bm{a}_t)\Big\}\\
&=\sum_{t=0}^{T-1} \sum_{ \bm{s}_t\in \mathcal{S}_t }  \frac{-\pi^{\bm{\sigma}^k}_t ( \bm{s}_t) }{[\Sigma^k( \bm{s}_t)]^2} \Big[ \sum_{ \bm{a}_t=1}^{|\mathcal{A}_t|} \sigma_t^k( \bm{s}_t, \bm{a}_t)[A_t^{\bm{\sigma}^k}( \bm{s}_t, \bm{a}_t)]^2\\
&-\sum_{\bm{a}_t=1}^{|\mathcal{A}_t|} \frac{\sigma^k(\bm{s}_t, \bm{a}_t)}{\Sigma^k(\bm{s}_t)} A_t^{\bm{\sigma}^k}(\bm{s}_t, \bm{a}_t) \sum_{\bm{b}_t=1}^{|\mathcal{A}_t|}\sigma_t^k( \bm{s}_t, \bm{b}_t)A_t^{\bm{\sigma}^k}( \bm{s}_t,  \bm{b}_t)\Big]\\
\end{split}
\end{eqnarray}
\normalsize

We can set  $\Sigma^0(\bm{s}_t)=1$ ($\forall \bm{s}_t \in \mathcal{S}_t$) for the initial  $\bm{\sigma}^0$  in the GBPI method, which is trivial.  Therefore, we have

\vspace{-6mm}
{\small
\begin{equation}
\begin{split}
&J(\bm{\sigma}^{k+1}; S_0)-J(\bm{\sigma}^{k}; S_0)\\
&\!=\!\sum_{t=0}^{T-1} \!\sum_{ \bm{s}_t\in \mathcal{S}_t } \!\sum_{\bm{a}_t=1}^{|\mathcal{A}_t|} \Big\{\frac{-\pi^{\bm{\sigma}^k}_t ( \bm{s}_t) }{[\Sigma^k( \bm{s}_t)]^3}\Big[\Sigma^k(\bm{s}_t)\gamma_t^k (\bm{s}_t, \bm{a}_t)A_t^{\bm{\sigma}^k}( \bm{s}_t, \bm{a}_t) \\
\end{split}
\end{equation}}
{\small 
\begin{equation*}
\begin{split}
&\quad \quad \!-\!\gamma_t^k( \bm{s}_t,  \bm{a}_t) \sum_{ \bm{a}_t=1}^{|\mathcal{A}_t|}\sigma_t^k( \bm{s}_t, \bm{a}_t)A_t^{\bm{\sigma}^k}(\bm{s}_t, \bm{a}_t)\Big] A_t^{\sigma^k}( \bm{s}_t, \bm{a}_t)\Big\}\\
&=\sum_{t=0}^{T-1} \sum_{\bm{s}_t\in \mathcal{S}_t }  \frac{-\pi^{\sigma^k}_t (\bm{s}_t) }{[\Sigma^k(\bm{s}_t)]^2}M(\bm{s}_t)\\
\end{split}
\end{equation*}}
where we have $M(\bm{s}_t)=\sum_{\bm{a}_t=1}^{|\mathcal{A}_t|} \sigma_t^k(\bm{s}_t, \bm{a}_t)[A_t^{\bm{\sigma}^k}(\bm{s}_t, \bm{a}_t)]^2-\sum_{\bm{a}_t=1}^{|\mathcal{A}_t|} \sigma_t^k(\bm{s}_t, \bm{a}_t) A_t^{\bm{\sigma}^k}(\bm{s}_t, \bm{a}_t )\sum_{\bm{b}_t=1}^{|\mathcal{A}_t|}  \sigma_t^k(\bm{s}_t, \bm{b}_t)A_t^{\bm{\sigma}^k}(\bm{s}_t, \bm{b}_t )$.

Since  we have $\sum_{\bm{a}_t=1}^{|\mathcal{A}_t|}\sigma_t^k(\bm{s}_t, \bm{a}_t)=\Sigma^k(\bm{s}_t) = 1$ ($\forall \bm{s}_t \in \mathcal{S}_t$) and the function $f(x)=x^2$ convex,  we  conclude  $M(\bm{s}_t)\geq 0$ ($\forall \bm{s}_t \in \mathcal{S}_t $) based on the properties of convex functions. Further, we have
\begin{equation} \label{perforemance}
\begin{split}
J(\bm{\sigma}^{k+1}; \bm{S}_0)\!-J(\bm{\sigma}^{k}; \bm{S}_0)\leq 0
\end{split}
 \end{equation}
The inequality \eqref{perforemance}  implies that the  performance  function $J(\bm{\sigma}^{k}; S_0)$ is  non-increasing w.r.t.  iteration $k$. 

As problem (\ref{Optimization Problem}) is bounded (i.e., the action space is bounded), we imply that the GBPI  method can converge to a local optima $\bar{\bm{\sigma}}$ with $k\rightarrow\infty$.  
The remainder illustrates that the method can converge to a global optima.

We assume that at least one global optimal  policy $\bm{\sigma}^*$  exists  for problem (\ref{Optimization Problem}). 
It  is straightforward that  $$J(\bm{\sigma}^{*}; S_0) \leq J(\overline{\bm{\sigma}}; S_0).$$
 
 We define a parameterized  random  policy as  $\bm{\sigma}^{r}=(1-r) \bar{\bm{\sigma}}+ r \bm{\sigma}^*$ ($\delta\in [0, 1]$). Since $\overline{\bm{\sigma}}$ is a local optima,  for any  $ r $ sufficiently small,  we have
 \begin{equation} 
 \begin{split}
 &J(\bm{\sigma}^{r}; \bm{S}_0) -J(\overline{\bm{\sigma}}; \bm{S}_0)\geq 0
 \end{split}
 \end{equation}  
 
 According to \eqref{(34)} and  $\Sigma(\bm{s}_t) = 1$ ($\forall \bm{s}_t \in \mathcal{S}_t$),  we have
\small
 \begin{equation} \label{(44)}
 \begin{split}
&J(\bm{\sigma}^{r}; S_0) -J(\overline{\bm{\sigma}}; S_0)\\
&=\sum_{t=0}^{T-1} \sum_{ \bm{s}_t\in \mathcal{S}_t} \Big[ \sum_{\bm{a}_t=1}^{|\mathcal{A}_t|} (\sigma_t^{\bm{r}}(\bm{s}_t, \bm{a}_t)-\bar{\sigma}_t(\bm{s}_t, \bm{a}_t))A_t^{\bar{\sigma}}(\bm{s}_t, \bm{a}_t)\Big]\\
&\!=\! r \sum_{t=0}^{T-1} \!\sum_{\bm{s}_t\in \mathcal{S}_t }\! \Big[ \sum_{\bm{a}_t=1}^{|\mathcal{A}_t|}  (\bm{\sigma}^{*}_t( \bm{s}_t, \bm{a}_t)-\bar{\sigma}_t( \bm{s}_t, \bm{a}_t))A_t^{\bar{\sigma}}(\bm{s}_t, \bm{a}_t)\Big] \geq 0\\
 \end{split}
 \end{equation}
 \normalsize
 
On the other hand, since $\bm{\sigma}^*$ is  global optima, we have 
 \small
  \begin{equation} \label{(45)}
 \begin{split}
 &J(\bm{\sigma}^{*}; \bm{S}_0) -J(\overline{\bm{\sigma}}; \bm{S}_0)\\
 =&\sum_{t=0}^{T-1} \sum_{\bm{s}_t\in \mathcal{S}_t } \Big[ \sum_{\bm{a}_t=1}^{|\mathcal{A}_t|}(\sigma_t^{*}(\bm{s}_t, \bm{a}_t)-\bar{\bm{\sigma}}_t(\bm{s}_t, \bm{a}_t))A_t^{\bar{\sigma}}(\bm{s}_t, \bm{a}_t)\Big]\\
 & \leq 0
 \end{split}
 \end{equation}
 \normalsize
By contrasting   (\ref{(44)}) and (\ref{(45)}), we conlude  $\bar{\bm{\sigma}}=\bm{\sigma}^*$,  otherwise they contradict with each other.   This implies the GBPI method can converge to the global optima of problem \eqref{Optimization Problem}. 
\end{proof}
\ifCLASSOPTIONcaptionsoff
  \newpage
\fi



%
\bibliographystyle{ieeetr}
\bibliography{reference}

%








\end{document}